\documentclass{cernyrep} 
\usepackage{texnames}
\usepackage[T1]{fontenc}
\RequirePackage{xspace}
\usepackage{relsize}
\def\babar{\mbox{\slshape B\kern-0.1em{\smaller A}\kern-0.1em B\kern-0.1em{\smaller A\kern-0.2em R}}\xspace}

\begin{document}
\title{Probability and Statistics for Particle Physicists}
 
\author{Jos\'{e} Ocariz}

\institute{Universit\'e Paris-Diderot and Laboratoire de physique nucl\'eaire et des hautes \'energies LPNHE  CERN-IN2P3, Paris, France}

\maketitle

\begin{abstract}
A pedagogical selection of topics in probability and statistics is presented.
Choice and emphasis are driven by the
author's personal experience, predominantly
in the context of physics analyses using experimental data from high-energy physics detectors.
\end{abstract}
 
\section{Introduction}
These notes are based on a series of three lectures on probability and statistics, given at the AEPSHEP 2012 physics school.
While most of the attendance was composed of PhD students, it also included Master students 
and young post-docs;
in consequence, a variable level of familiarity with the topics discussed was implicit. For consistency,
the scope of the lectures spanned from very general concepts up to more advanced, recent developments.
The first lecture reviewed basic concepts in probability and statistics; the second lecture
focussed on maximum likelihood and multivariate techniques for statistical analysis of experimental
data; the third and last lecture covered topics on 
hypothesis-testing and interval estimation. Whenever possible, the  notation aligns with  common usage
in experimental high-energy physics (HEP), and the discussion is illustrated with examples
related to recent physics results, mostly from the $B$-factories and the LHC experiments.

\section{Basic concepts in probability and statistics}
Mathematical probability is an abstract axiomatic concept,
and  probability theory is the conceptual
framework to assess the  knowledge of random processes. 
A detailed discussion of its development and formalism lies outside
the scope of these notes. Other than standard classic books, like~\cite{bib:Kendall}, there
are excellent references available,  often written by (high-energy) physicists, 
and well-suited for the needs of physicists.
A non-comprehensive list
includes~\cite{bib:Barlow,bib:Cowan,bib:James,bib:Lyons}, and can guide the reader into more advanced topics.
The sections on statistics and probability in the PDG~\cite{bib:PDG} are also a useful reference; often also,
the large experimental collaborations have (internal) forums and working groups, with many useful links and
references. 
\subsection{Random processes}
For a process to be  called random, two main conditions are required:
	its outcome cannot be predicted with complete certainty, and 
	if the process is repeated under the very same conditions, the new resulting outcomes can be different each time.	 
In the context of experimental particle physics,
		such an  outcome  could be 
		``a collision'', or ``a decay''.
In practice, the sources of uncertainty leading to random processes can be 
\begin{itemize}
	\item	due to reducible measurement errors, i.e. practical limitations that can in principle be 
		overcome by means of higher-performance
		instruments or improved control of experimental conditions;
	\item	due to quasi-irreducible random measurement errors,  i.e. thermal effects;
	\item	fundamental, if the underlying physics is intrinsically uncertain, i.e. quantum mechanics
		is not a deterministic theory.
\end{itemize}
Obviously in particle physics, all three kinds of uncertainties are at play. A key feature
of collider  physics is that events
resulting from particle collisions
are independent of each other, and provide a quasi-perfect laboratory of quantum-mechanical probability processes.
Similarly, unstable particles produced in HEP experiments obey quantum-mechanical decay probabilities.
\subsection{Mathematical probability}
	Let $\Omega$ be the total universe of possible outcomes of a random process,
and let $X,Y\dotsc$ be elements (or realizations)
of $\Omega$; a set of such realizations is called a sample. A probability function ${\cal P}$ is defined as a map onto the real numbers:
\begin{eqnarray}
	{\cal P} : \left\{ \Omega \right\} & \rightarrow & \left[ 0 : 1 \right] \ ,
\nonumber \\
X & \rightarrow & {\cal P}(X) \ . 
\end{eqnarray}
This mapping  must satisfy the following axioms:
\begin{eqnarray}
	{\cal P}(\Omega) & = & 1  \   , 
\nonumber \\
	{\rm if} \ X \cap Y & = & \oslash  \  ,  \ {\rm then} \ {\cal P}( X \cup Y ) = {\cal P}( X ) + {\cal P}( Y ) \ ,
\end{eqnarray}
from which various useful properties can be easily derived, i.e.
\begin{eqnarray}
	{\cal P}(\overline{X}) & = & 1 - {\cal P}(X) \ ,
\nonumber \\
	{\cal P}(X\cup \overline{X}) & = & 1 \ ,
\nonumber \\	
	{\cal P}(\oslash) & = & 1 - {\cal P}(\Omega) = 0 \ ,
\nonumber \\
	{\cal P}(X\cup Y) & = & {\cal P}(X) + {\cal P}(Y) - {\cal P}(X\cap Y) \ ,
\end{eqnarray}
(where $\overline{X}$ is the complement of $X$).
Two elements $X$ and $Y$ are said to the independent (that is, their realizations
are not linked in any way) if
\begin{eqnarray}
\label{eq:independence}
{\cal P}(X\cap Y) = {\cal P}(X) {\cal P}(Y).
\end{eqnarray}
\subsubsection{Conditional probability and Bayes' theorem}
Conditional probability
${\cal P}(X\mid Y)$ is defined as the probability of $X$, given $Y$. The simplest example of conditional probability
is for independent outcomes:
from the definition of independence in Eq.~(\ref{eq:independence}), it follows that if $X$ and $Y$ are actually independent, the condition
\begin{eqnarray}
{\cal P}(X\mid Y) = {\cal P}(X) 
\end{eqnarray}
is satisfied. The general case is given by
Bayes' theorem: in view of the relation ${\cal P}(X\cap Y)={\cal P}(Y\cap X)$, it follows that
\begin{eqnarray}
\label{eq:bayes}
{\cal P}( X \mid Y ) = \frac{{\cal P}( Y \mid X ){\cal P}(X)}{{\cal P}(Y)} \ .
\end{eqnarray}
An useful corollary follows as consequence of Bayes' theorem: if $\Omega$ can be 
divided into a number of disjoint subsets $X_i$ 
(this division process is called a partition), then
\begin{eqnarray}
{\cal P}( X \mid Y ) = \frac{{\cal P}( Y \mid X ){\cal P}(X)}{\sum_i{\cal P}(Y \mid X_i){\cal P}(X_i)} \ .
\end{eqnarray}
\subsubsection{The probability density function}
\begin{figure}[ht]
\begin{center}
\includegraphics[width=7.5cm]{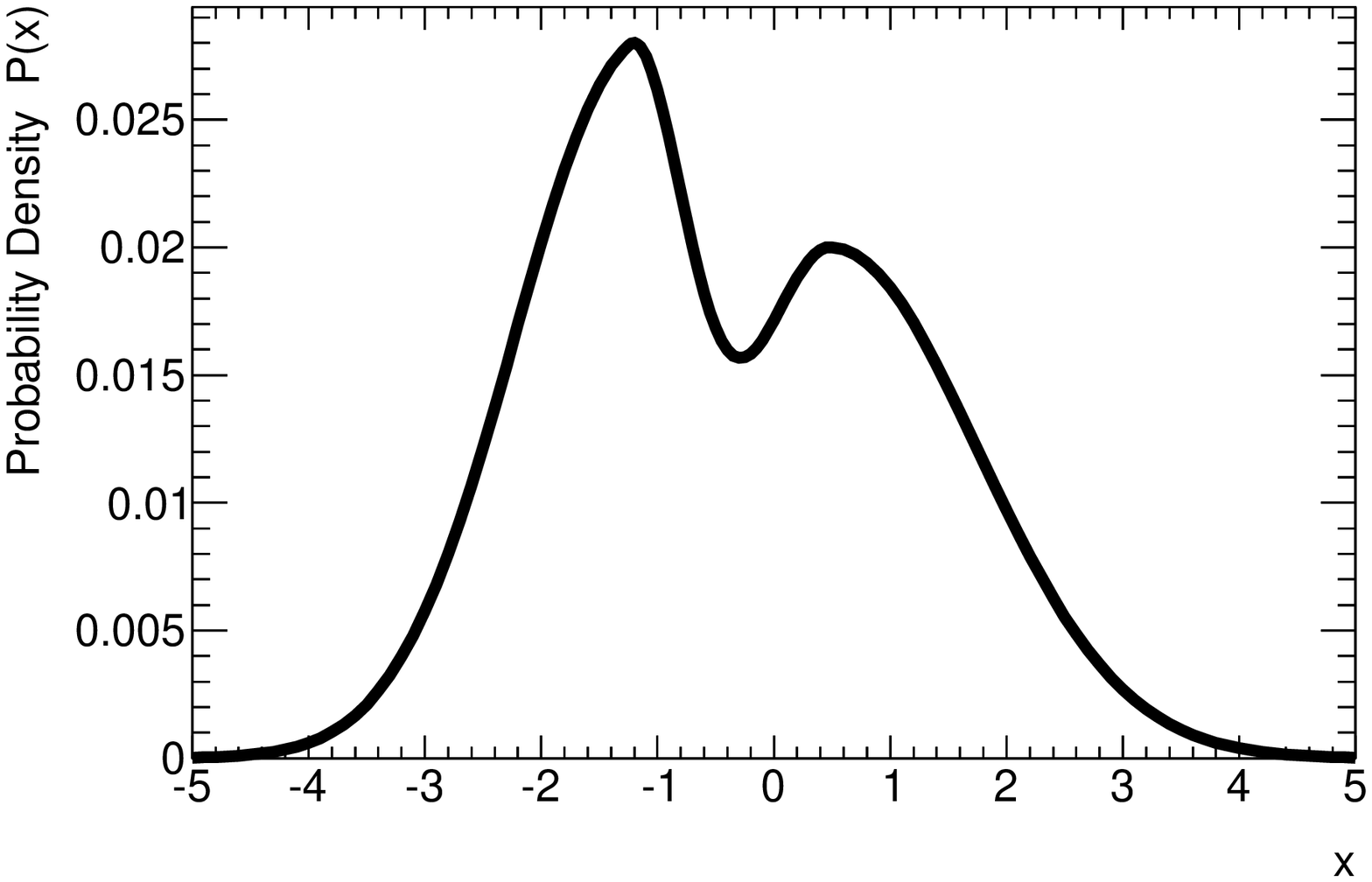}
\includegraphics[width=7.5cm]{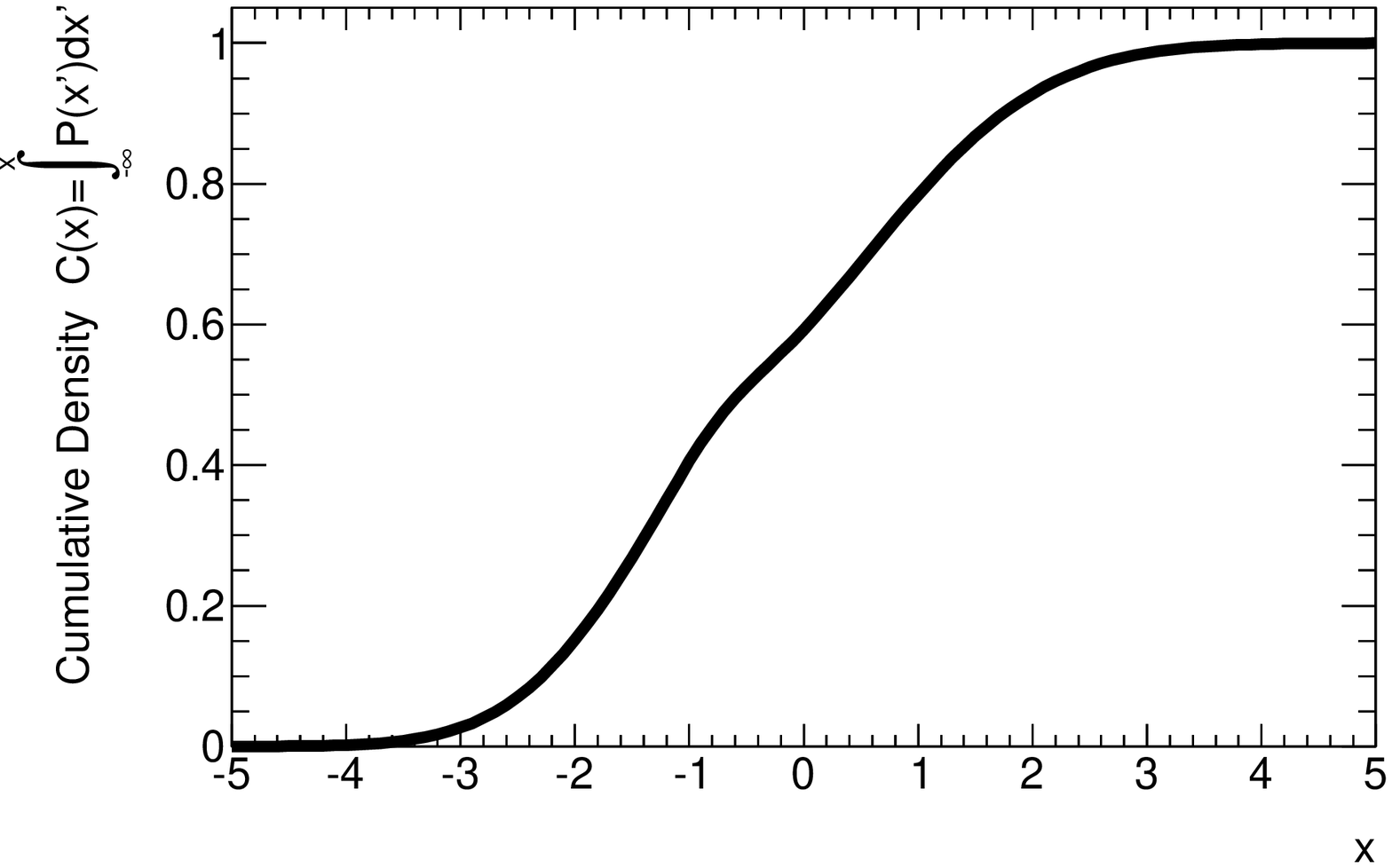}
\caption{Left: a probability density function (PDF) for a variable $x$; the PDF is assumed to have negligible 
values outside of the
plotted range. Right: the corresponding cumulative density function (CDF), plotted in the same range.}
\label{fig:pdfCanvas}
\end{center}
\end{figure}

In the context of these lectures, the relevant scenario is when the outcome of a random process can be stated
in numerical form (i.e. it corresponds to a measurement): then  to each element $X$ (which for HEP-oriented
notation purposes, it is preferable to design as an event) corresponds a variable $x$
(that can be real or integer).
For continuous $x$, its probability density function (PDF) $P(x)$ is defined as
\begin{eqnarray}
{\cal P}(X \ {\rm found} \ {\rm in} \ \left[x,x+dx\right] ) \ = \ P(x)dx \ ,
\end{eqnarray}
where $P(x)$ is positive-defined for all values of $x$, and satisfies the normalization condition
\begin{eqnarray}\label{eq:normalization}
\int_{-\infty}^{+\infty} dx^\prime P(x^\prime) \ = \ 1 \ .
\end{eqnarray}
For a discrete $x_i$, the above definition can be adapted in a  straightworward way: 
\begin{eqnarray}
{\cal P}(X \ {\rm found} \ {\rm in} \ x_i) \ = \ p_i \ , 
\nonumber \\
{\rm with} \ \sum_j p_j = 1 \ {\rm and} \ p_k \geq 0 \ \forall k \ . 
\end{eqnarray}
Finite probabilities are obtained by integration over a non-infinitesimal
range. It is sometimes convenient to refer to the cumulative density function (CDF) :
\begin{eqnarray}
C(x) \ = \ \int_{-\infty}^{x} dx^\prime P(x^\prime) \ ,
\end{eqnarray}
so that finite probabilities can be obtained by evaluating the CDF on the boundaries of the range of interest :
\begin{eqnarray}
{\cal P}(a<X<b) \ = \ C(b) - C(a) \ = \ \int_a^b dx^\prime P(x^\prime) \ .
\end{eqnarray}
Other than the conditions of normalization Eq.~(\ref{eq:normalization}) and positive-defined (or more precisely, 
to have a compact support, which
implies that the PDF must become vanishingly small outside of some finite boundary)
the PDFs can be arbitrary otherwise, and exhibit one or several local maxima or local minima.
In contrast, the CDF is a monotonically increasing function of $x$, as shown on
Figure~\ref{fig:pdfCanvas}, where a generic PDF and its corresponding CDF are represented.
\subsubsection{Multidimensional PDFs}
\begin{figure}[ht]
\begin{center}
\includegraphics[width=9.5cm]{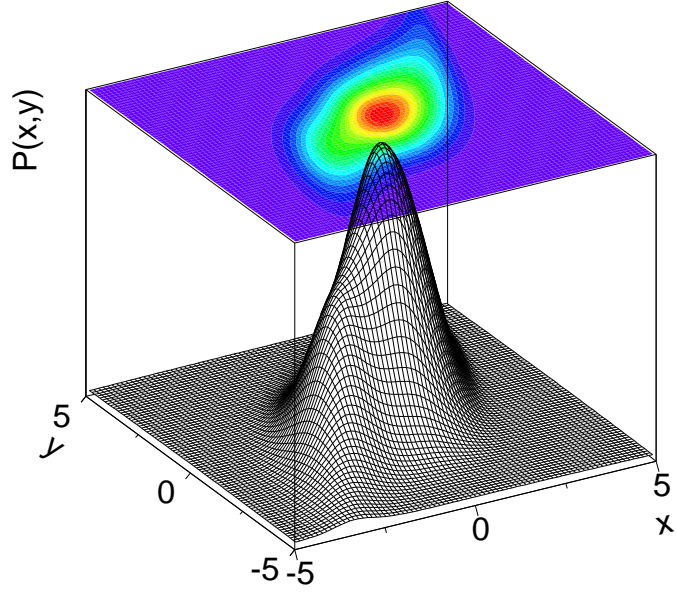}
\caption{A two-dimensional probability density function (PDF) for 
non-independent variables $x$ and $y$: the pattern implies that in general, the probability densities are larger when 
$x$ and $y$ are both large or both small (i.e. they are positively correlated), and thus the variables are not independent. The PDF is assumed to have negligible 
values outside of the
plotted two-dimensional range.}
\label{fig:pdf2DCanvas}
\end{center}
\end{figure}

When more than on random number is produced as outcome in a same event, it is convenient to 
introduce a 
$n$-dimensional set of random elements $\vec{X}=\left\{X_1,X_2,\dotsc,X_n\right\}$, together with
its corresponding set of random variables $\vec{x}=\left\{x_1,x_2,\dotsc,x_n\right\}$
and their  multidimensional PDF:
\begin{eqnarray}
\label{eq:multidimensionalPDF}
P(\vec{x})d\vec{x} \ = \ P(x_1,x_2,\dotsc,x_n)dx_1dx_2\dotsc dx_n \ .
\end{eqnarray}
Lower-dimensional PDFs can be derived from Eq.~(\ref{eq:multidimensionalPDF}); for instance, when 
one specific  variable $x=x_j$ (for fixed $j$, with $1\leq j\leq n$) is of particlar relevance,
its one-dimensional marginal probability density $P_X(x)$  is extracted by 
integrating $P(\vec{x})$  over the remaining $n-1$ dimensions (excluding the $j$-th):
\begin{eqnarray}
P_X(x)dx = dx\int_{-\infty}^{+\infty}dx_1\dotsc \int_{-\infty}^{+\infty}dx_{j-1} \int_{-\infty}^{+\infty}dx_{j+1} \dotsc \int_{-\infty}^{+\infty}dy_{n-1} \ .
\end{eqnarray}
Without loss of generality, the discussion can be restricted to the two-dimensional case, with 
random elements $X$ and $Y$ and random variables
$\vec{X}=\left\{x,y\right\}$. The finite probability in a rectangular two-dimensional range is 
\begin{eqnarray}
{\cal P}(a<X<b \ {\rm and} \ c<Y<d) \ = \ \int_a^b dx \int_c^d dy P(x,y) \ .
\end{eqnarray}
For a fixed value
of $Y$, the conditional density function for $X$ is given by 
\begin{eqnarray}
P(x \mid y) = \frac{P(x,y)}{\int dyP(x,y)} = \frac{P(x,y)}{P_Y(y)} \ .
\end{eqnarray}
As already mentioned,  the relation $P(x,y)=P_X(x)\cdot P_Y(y)$ holds only if $X$ and $Y$ are independent; for instance,
the two-dimensional density function in Figure~\ref{fig:pdf2DCanvas} is an example of non-independent variables, for which
$P(x,y) \neq P_X(x)\cdot P_Y(y)$. 
\section{Parametric PDFs and parameter estimation}\label{sec:paramestimation}
The description of a random process via density functions is called a model.
Loosely speaking, a parametric model assumes that its PDFs can be completely
described using a finite number of parameters~\footnote{This requirement needs not to be satisfied; PDFs can also
be non-parametric (which is equivalent to assume that an infinite number of parameters is needed to describe them),
or they can be a  mixture of both types.}. A straightforward implementation of a parametric PDF is when its parameters
are analytical arguments of the density function; the notation
$P\left(x,y,\dotsc;\theta_1,\theta_2,\dotsc\right)$ indicates the functional dependence of the PDF (also called its shape) 
in terms of variables
$x_1,y_2,\dotsc$ and parameters $\theta_1,\theta_2,\dotsc$. 
\subsection{Expectation values}
\begin{figure}[ht]
\begin{center}
\includegraphics[width=3.75cm]{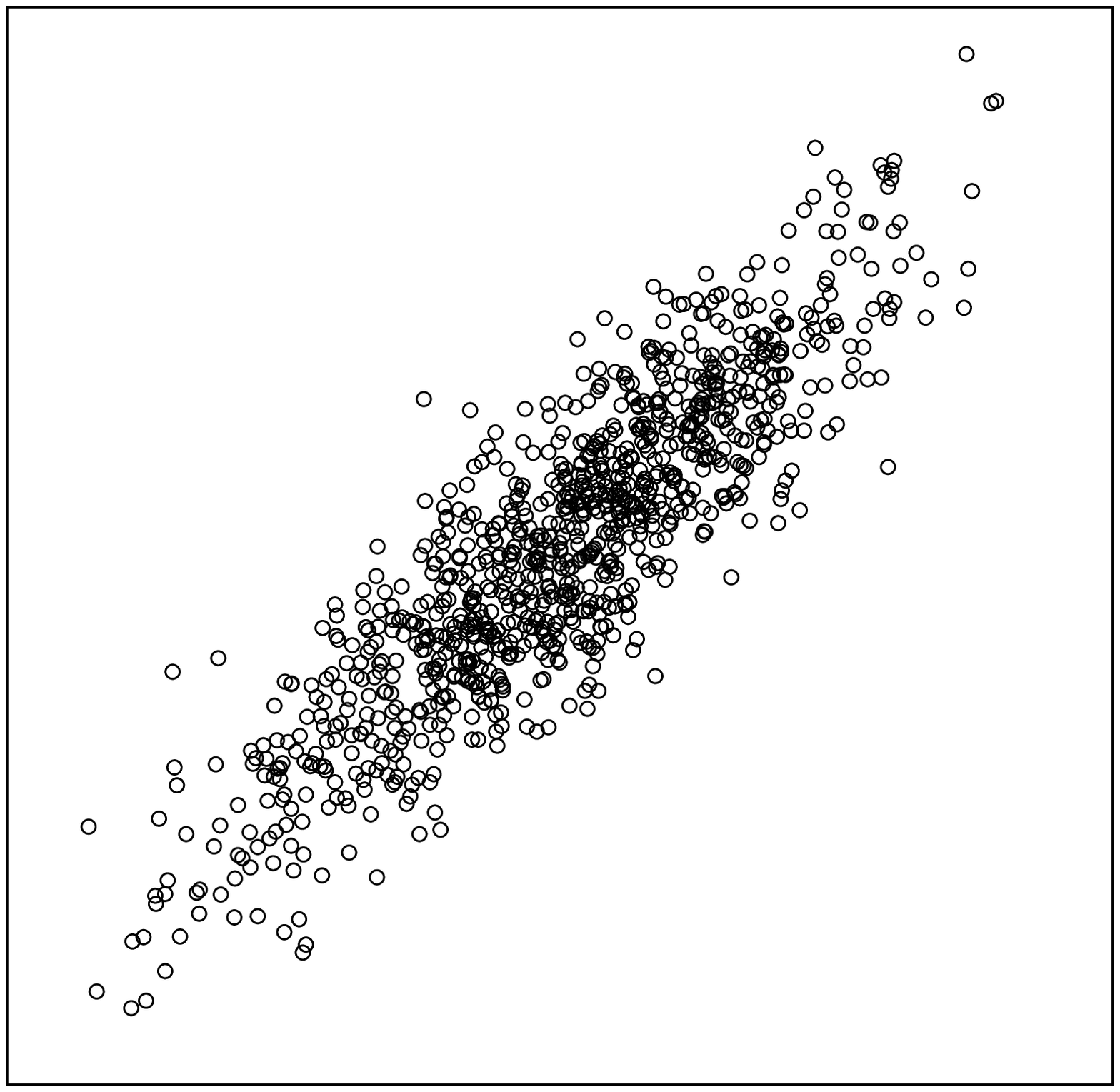}
\includegraphics[width=3.75cm]{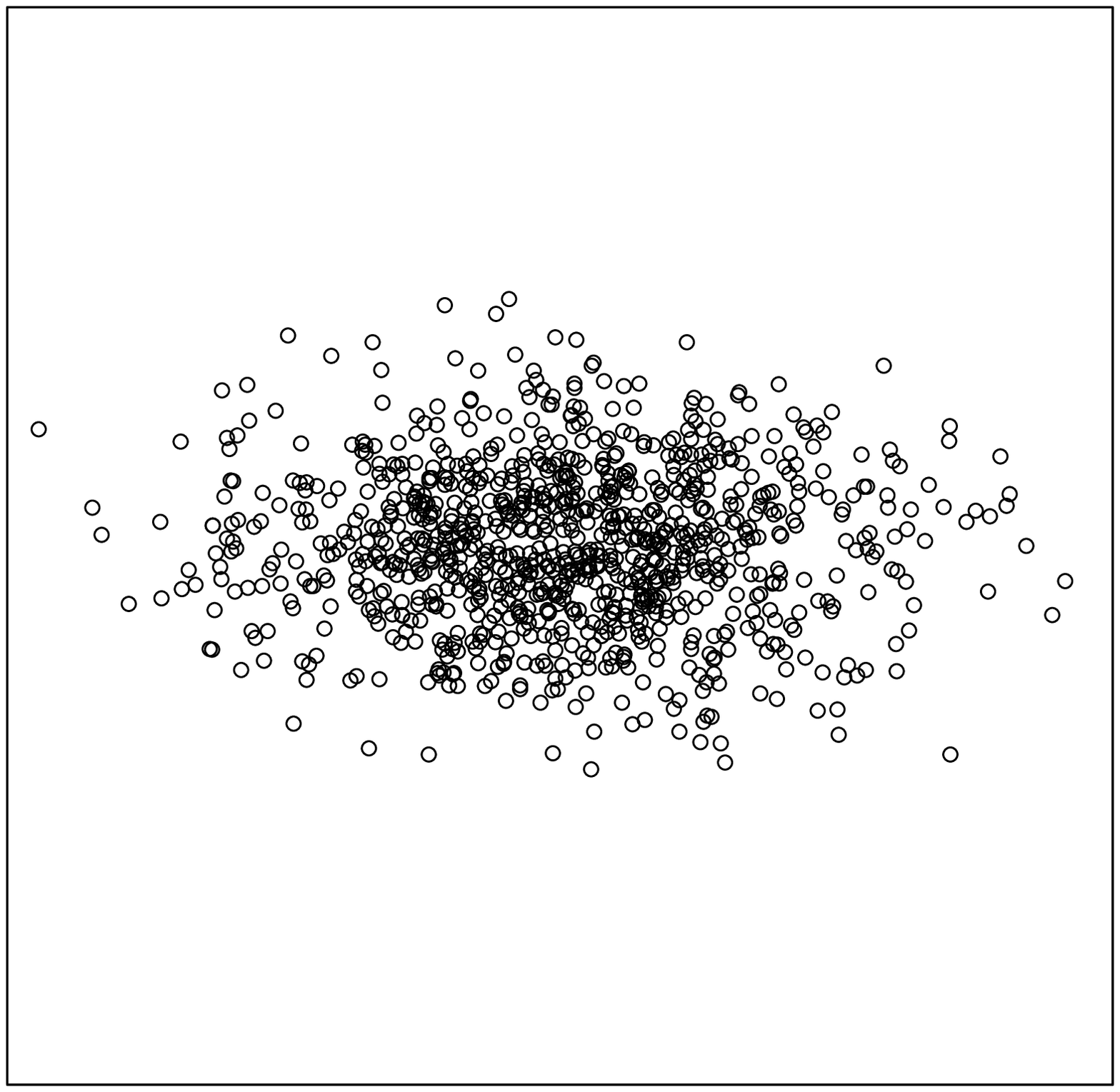}
\includegraphics[width=3.75cm]{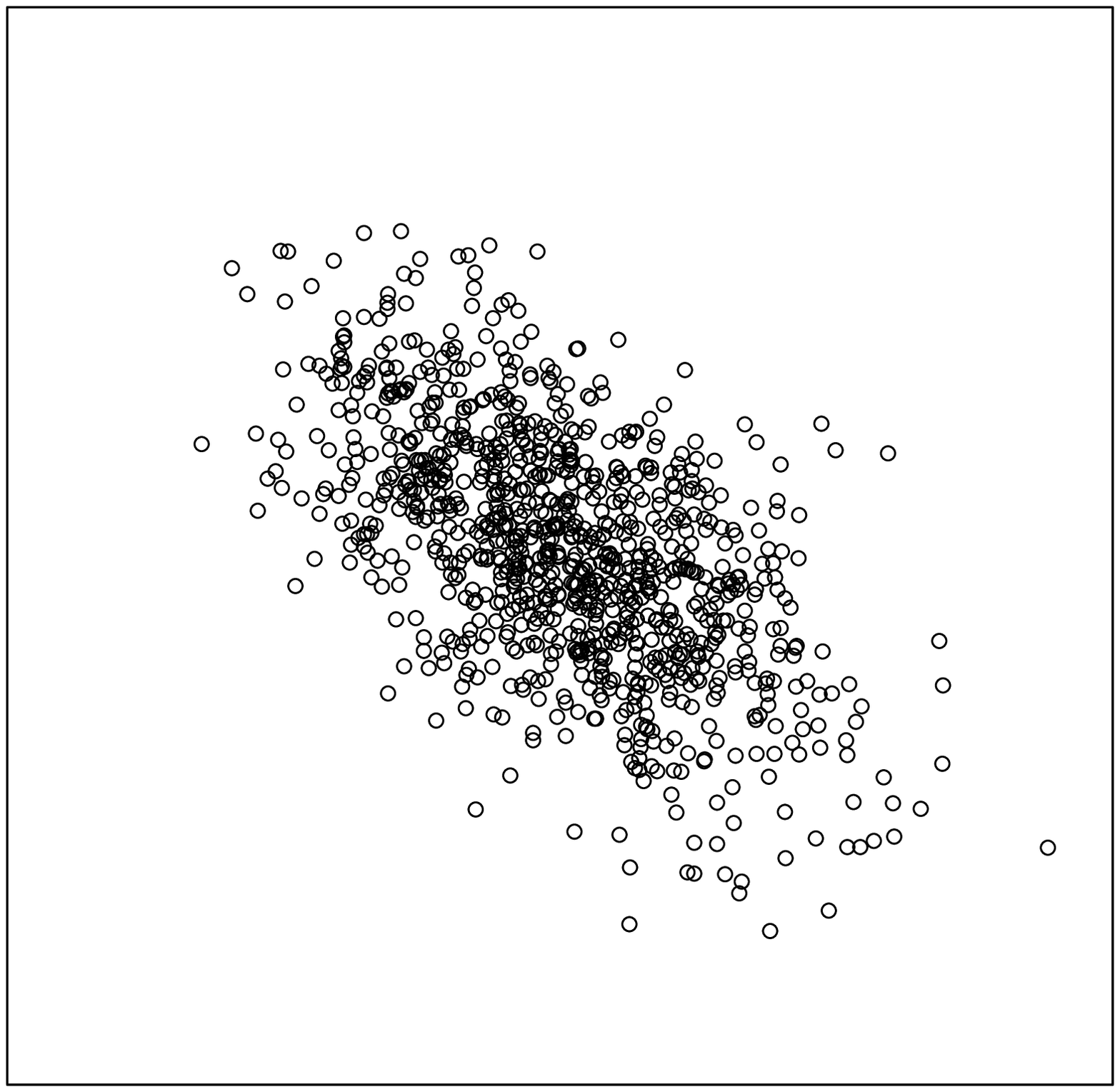}
\includegraphics[width=3.75cm]{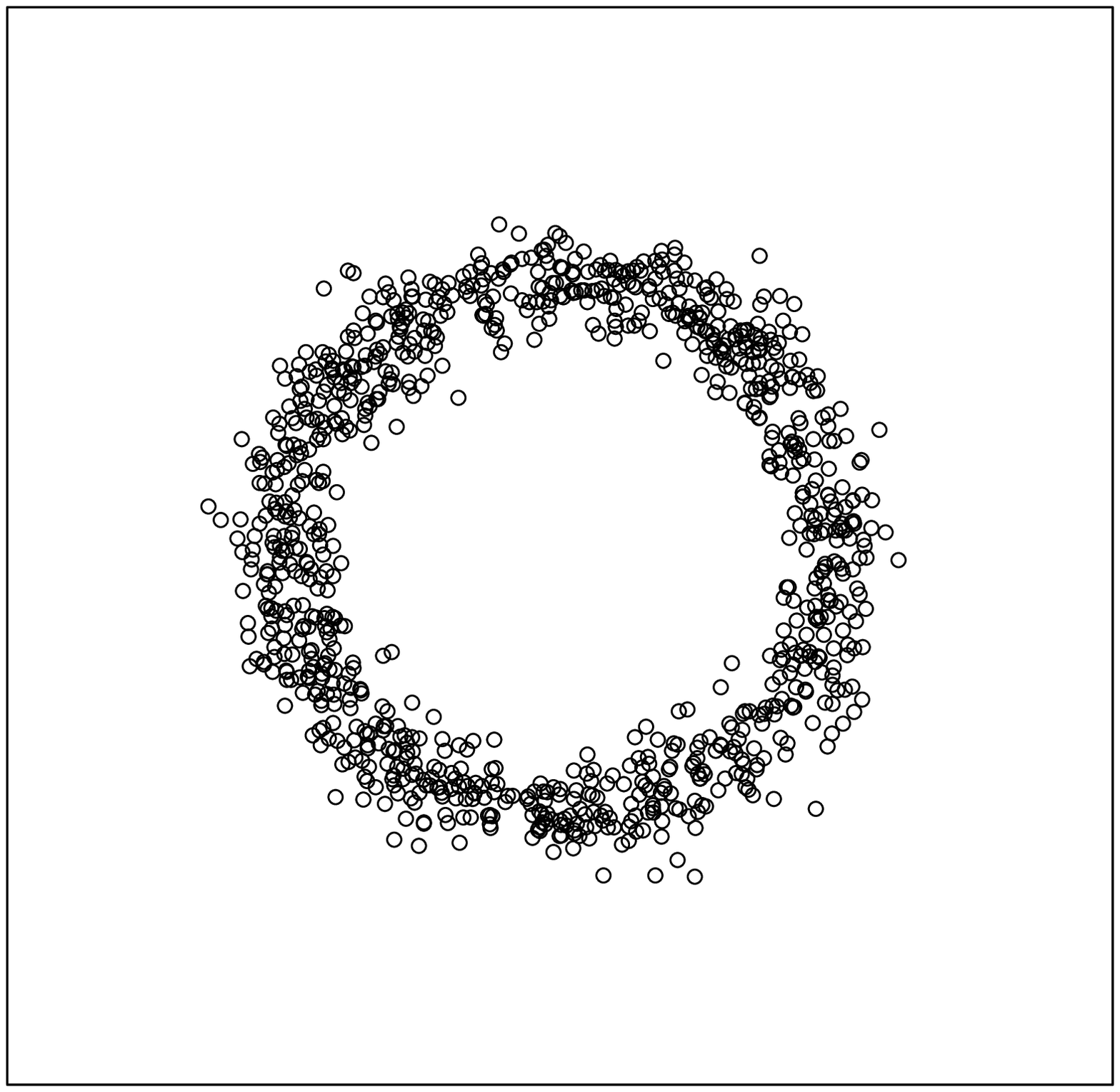}
\caption{Examples of two-dimensional probability density functions, illustrating four typical correlation
patterns. From left ro right, the figures show two variables which exhibit:
 a large, positive linear correlation (with $\rho=+0.9$);
no correlation, with a zero linear correlation coefficient;
a slightly milder, negative correlation (with $\rho=-0.5$); 
and a more complex correlation pattern, with variables
very strongly correlated, but in such a way that the linear correlation coefficient is zero.}
\label{fig:corrCanvas}
\end{center}
\end{figure}

Consider a random variable $X$ with PDF $P(x)$. For a generic function $f(x)$, its expectation value
$E[f]$ is the PDF-weighted average over the $x$ range : 
\begin{eqnarray}
E[f] \ = \ \int dx P(x) f(x) \ .
\end{eqnarray}
Being often used, some common expectation values have their own names. For one-dimensional PDFs, the
mean value and variance are defined as
\begin{eqnarray}
{\rm Mean \ value} & : \ \mu \ = & \ E[x] \ = \ \int dx P(x) x \ ,
\\
{\rm Variance} & : \ \sigma^2 \ = & \ V[x] \ = \ E[x^2] - \mu^2 \ = \ E[(x-\mu)^2] \ ;
\end{eqnarray}
for multidimensional PDFs, the covariance $C_{ij}=C(x_i,x_j)$ and the dimensionless linear correlation
coefficient $\rho_{ij}$ are defined as:
\begin{eqnarray}
{\rm Covariance} & : \ C_{ij} \ = & \ E[x_ix_j] -\mu_i\mu_j \ = \  E[(x_i-\mu_i)(x_j-\mu_j)] \ ,
\\
{\rm Linear} \ {\rm  correlation} & : \ \rho_{ij} \ = & \ \frac{C_{ij}}{\sigma_i\sigma_j} \ .
\end{eqnarray}
By construction, linear correlation coefficients have values in the $-1\leq\rho_{ij}\leq 1$ interval. 
The sign of the $\rho$ coefficient indicates the dominant trend in the $(x_j;x_j)$ pattern: for positive
correlation, the probability density is larger when $x_i$ and $x_j$ are both small or large, while
a negative correlation indicates that large values of $x_i$ are preferred when small values of $x_j$ are realized (and viceversa).
When the random variables $X_i$ and $X_j$ are independent, that is $P(x_i,x_j)=P_{X_i}(x_i)P_{X_j}(x_j)$, one has
\begin{eqnarray}
E[x_ix_j] \ = \ \int\int dx_idx_jP(x_i,x_j) x_ix_j \ = \ \mu_i\mu_j \ ,
\end{eqnarray}
and thus $\rho_{ij}=0$: independent variables have a zero linear correlation coefficient. Note
that the converse needs not be true: non-linear correlation patterns among non-independent
variables may ``conspire'' and yield null values of the
linear correlation coefficient, for instance if negative and positive correlation patterns in different
regions of the $(x_i,x_j)$ plane cancel out. Figure~\ref{fig:corrCanvas} shows examples of two-dimensional
samples, illustrating a few representative correlation patterns among their variables.
\subsection{Shape characterisation}\label{sec:shape}
In practice, the  true probability density function may not be known,
and the accessible information can only be extracted from a finite-size sample
(say consisting of $N$ events), which is assumed to have been originated from an unknown PDF.
Assuming that this underlying PDF is parametric, a procedure to estimate its functional dependence
and the values of its parameters is called a characterisation of its shape.
Now, only a finite number of expectation values can be estimated from 
a finite-size sample.
Therefore, when choosing the set of parameters to be estimated, each should provide information as useful and complementary as possible;
such procedure, despite being intrinsically incomplete, can nevertheless prove quite powerful.

The procedure of shape characterisation is first illustrated  with the one-dimensional case of a single random variable $x$.
Consider the empirical average $\overline{x}$  (also called sample mean)
\begin{eqnarray}
\overline{x} \ = \ \frac{1}{N}\sum_{i=1}^{N}x_i \ .
\end{eqnarray}
As shown later in Sec.~\ref{sec:musigma}, $\overline{x}$  is a good estimator of the mean value $\mu$ of the underlying distribution $P(x)$.
In the same spirit, the quadratic sample mean (or
root-mean-square RMS), 
\begin{eqnarray}
{\rm RMS} \ = \ \sqrt{\overline{x^2}-\left(\overline{x}\right)^2} \ ,
\end{eqnarray}
is a reasonable estimator of its variance $\sigma^2$ (an improved
estimator of variance can be easily derived from the RMS,  as discussed later in Sec.~\ref{sec:musigma}).
Intuitively speaking, these two estimators together provide complementary information on the
``location'' and ``spread'' of the region with highest event density in $x$. 

The previous approach can be discussed in a more systematic way, by means of the 
characteristic function, which is a transformation from 
the $x$-dependence of the PDF $P(x)$ onto a $k$-dependence
of $C[k]$, defined as
\begin{eqnarray}
C[k] \ = \ E\left[e^{ik\frac{x-\mu}{\sigma}}\right] \ = \ \sum_j \frac{(ik)^j}{j!}\mu_j \ .
\end{eqnarray}
The coefficients $\mu$ of the expansion are called reduced moments;
by construction, the first moments are $\mu_1=0$ and $\mu_2=1$; these
values indicate that in terms of the rescaled variable $x^\prime=(x-\mu)/\sigma$, the PDF has been shifted to have zero mean
and scaled to have unity variance.  

\begin{figure}[ht]
\begin{center}
\includegraphics[width=7.5cm]{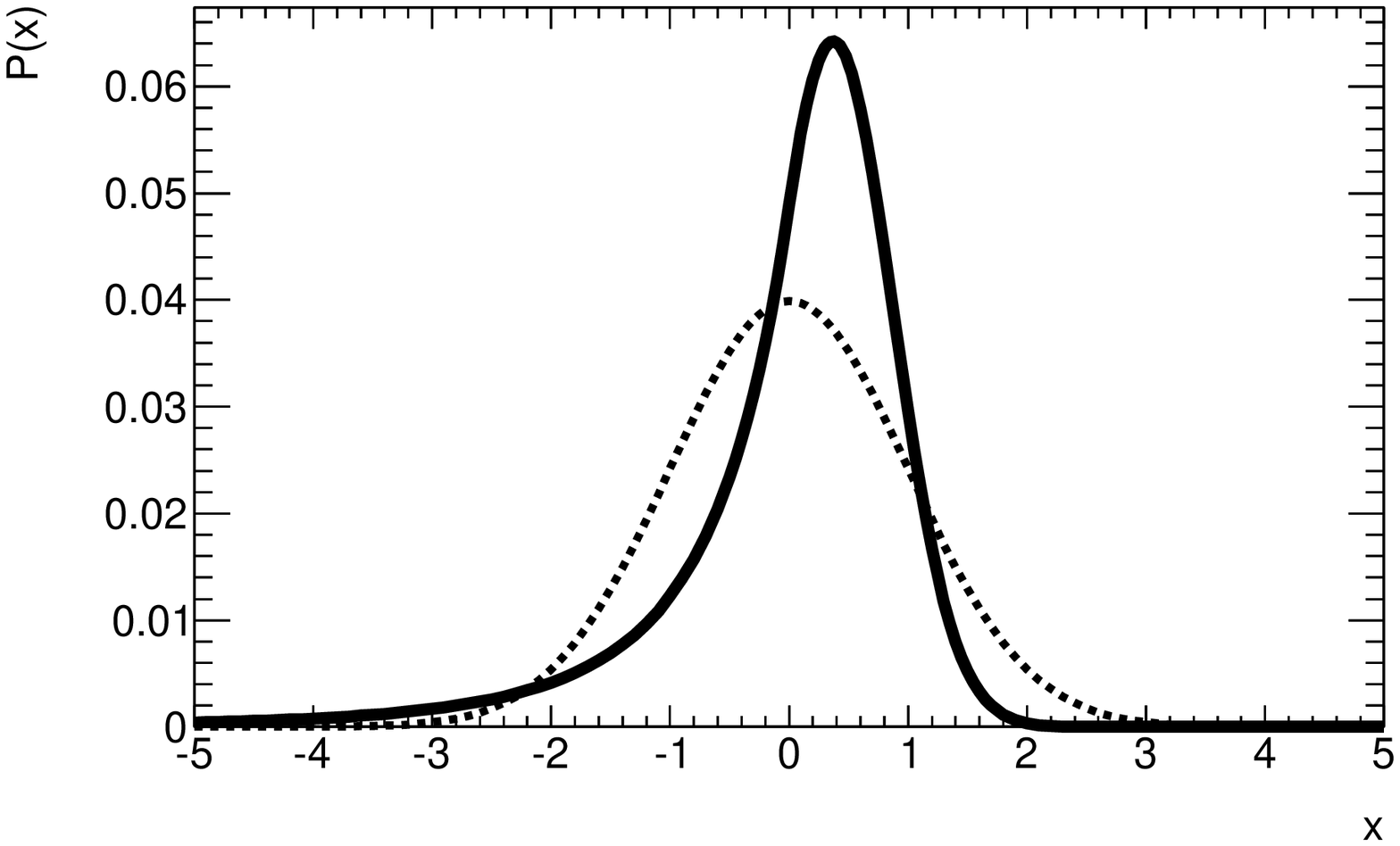}
\includegraphics[width=7.5cm]{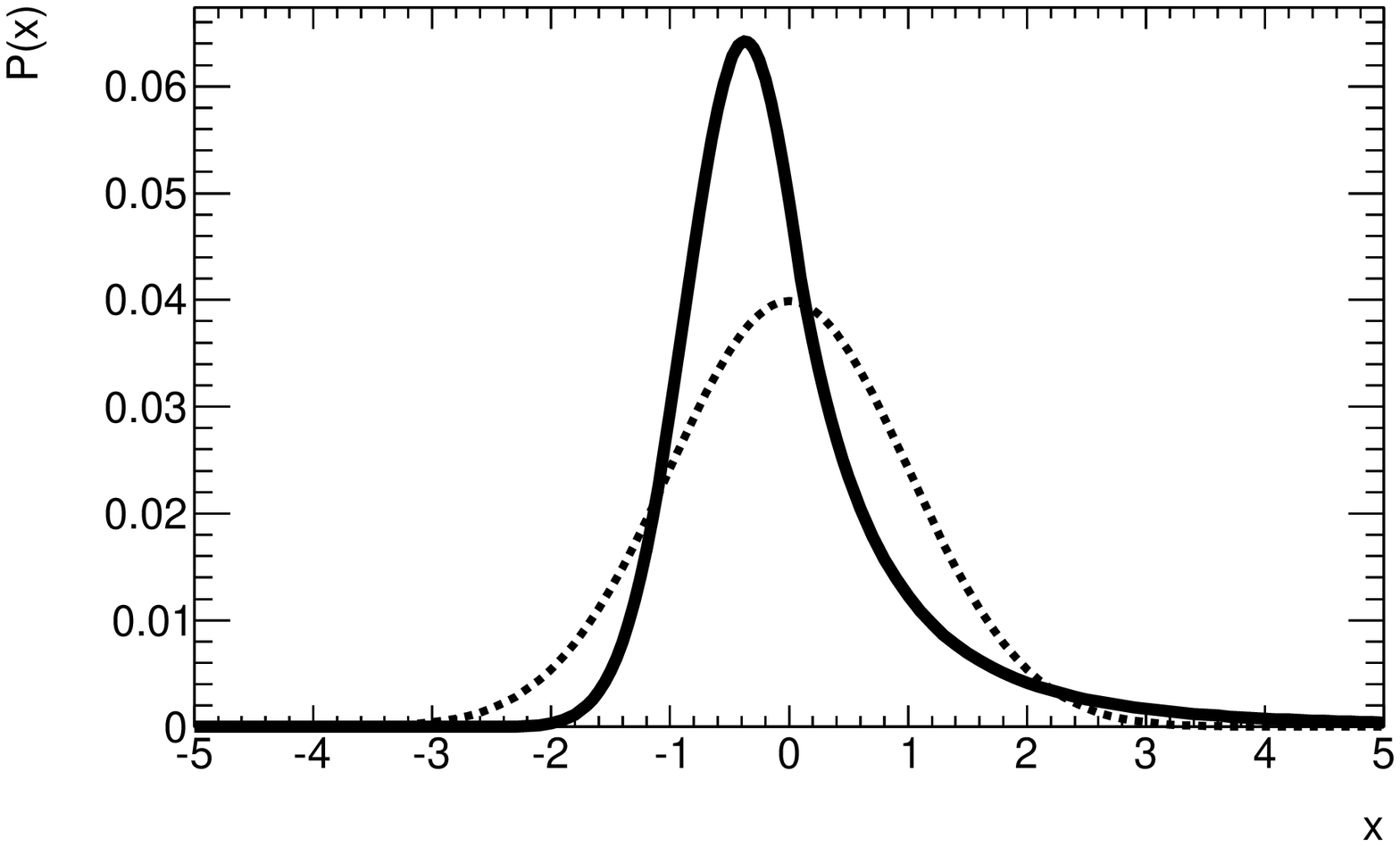}
\includegraphics[width=7.5cm]{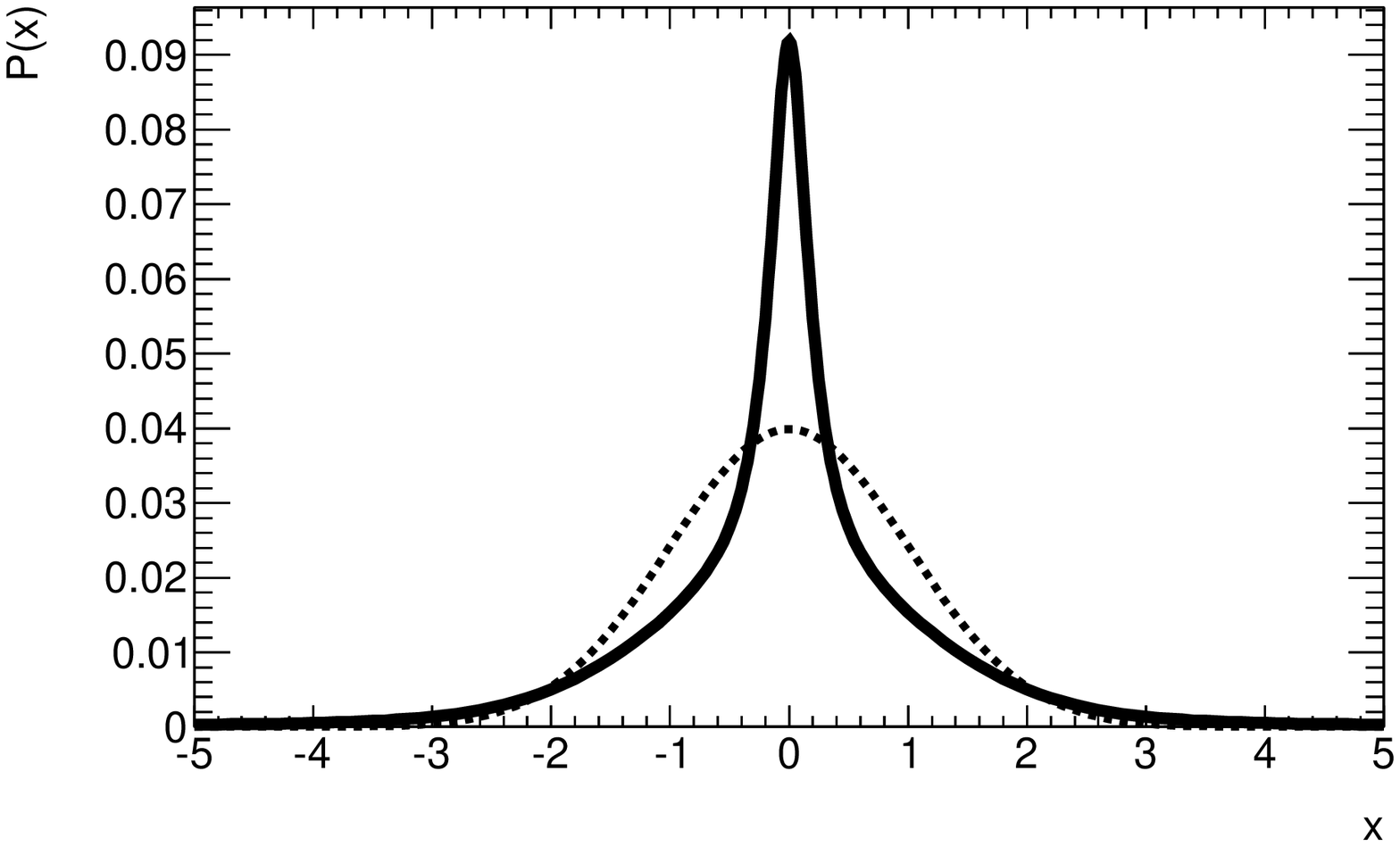}
\includegraphics[width=7.5cm]{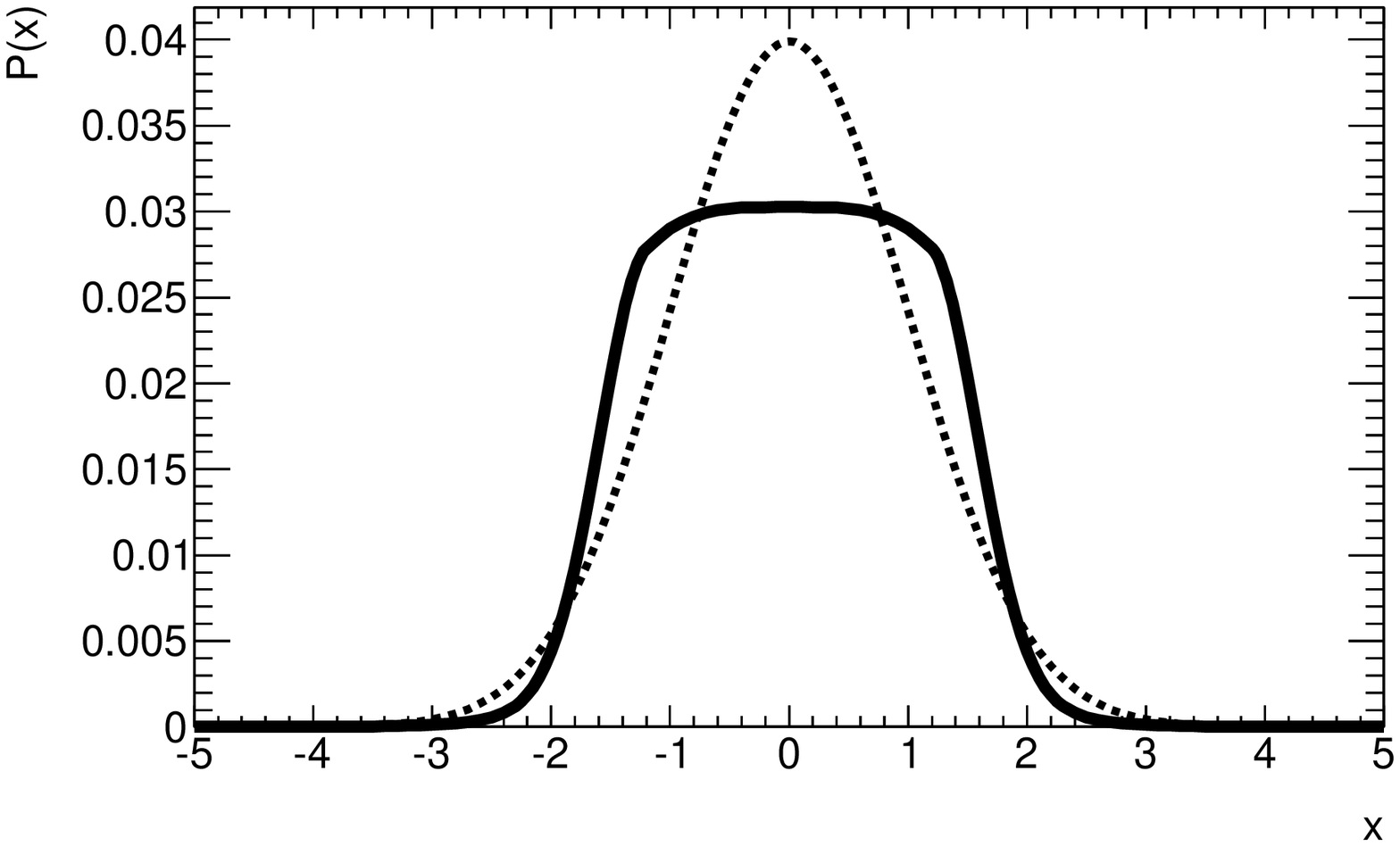}
\caption{Examples of probability density functions, illustrating the role of momenta in the characterization of
PDF shapes.  For all PDFs plotted, their mean values are 0 and their variances are 1. 
Top plots : the PDF on the left (resp. right) exhibits an asymmetric, non-gaussian tail to the left (resp. right) of its peak, and thus 
has a positive (resp. negative) skewness. Bottm plots: the PDF on the left (resp. right) shows a narrow peak and long-range tails
(resp. broad core and short-range tails), and has a large (resp. small)
kurtosis. In each plot, a Gaussian PDF (dashed line) with mean value 0 and variance 1, is also overlaid.}
\label{fig:kurtosisEtal}
\end{center}
\end{figure}

In principle, the larger the number of momenta $\mu_j$ that are estimated, the more detailed is the
characterisation of the PDF shape.
Among
higher-order moments, the third and fourth have specific names, and their values can be interpreted,
in terms of shape, in a relatively straightforward manner.
The third moment is called skewness: a symmetric distribution has zero skewness,
and a negative (positive) value indicates a larger spread to the left (right) of its median. 
The fourth moment is called kurtosis; it is a positive-defined quantity, that (roughly speaking) 
can be related to the "peakedness" of a 
distribution: a large value indicates a sharp peak and long-range tails (such a distribution is sometimes
called leptokurtic), while a smaller value reflects 
a broad central peak with short-range
tails (a so-called platykurtic distribution). 
Figure~\ref{fig:kurtosisEtal} shows a few distributions, chosen to illustrate the relation of momenta and shapes.
\subsection{Parameter estimation}\label{sec:parameterestimation}
The characterization of the shape of a PDF
through a sequential estimation of shape parameters discussed in Sec.~\ref{sec:shape}, aimed at a qualitative introduction to 
the concept of parameter estimation
(also called point estimation in the litterature). A more general approach is now discussed in this paragraph.

Consider a $n$-dimensional, $k$-parametric PDF,
\begin{eqnarray}
P( \ x_1,x_2,\dotsc ,x_n \ ; \ \theta_1,\theta_2,\dotsc ,\theta_k \ ) \ ,
\end{eqnarray}  
for which the values $\theta_1,\dotsc,\theta_k$ are to be estimated on a finite-sized sample, by means of a set of
estimators denoted $\hat{\theta_1},\dotsc,\hat{\theta_k}$.
The estimators themselves 
are random variables, with their own mean values and variances: their values differ when estimated on different samples.
The estimators should satisfy two key properties:
to be consistent and unbaised.
Consistency ensures that, in the infinite-sized sample limit, the estimator converges to the true parameter value; 
absence of bias
ensures that the expectation value of the estimator is, for all sample sizes, the true parameter value. 
A biased, but consistent estimator (also called asymptotically unbiased) is such that the bias 
decreases with increasing sample size.
Additional criteria can be used to characterize the quality and performance of estimators; two often mentioned are 
\begin{itemize}
	\item	efficiency: an estimator with small variance is said to be more efficient than one with larger variance;
	\item	robustness: this criterion characterizes the  sensitivity of an estimator to uncertainties in the underlying PDF. 
		For example, the mean value is robust against
		uncertainties on even-order moments, but is less robust with respect to  changes on odd-order ones.
\end{itemize}
Note that these criteria may sometimes be mutually conflicting; for practical reasons,  it may be preferable to use an efficient, 
biased estimator
to an unbiased, but poorly convergent one.

\subsubsection{The classical examples: mean value and variance}\label{sec:musigma}  
The convergence and bias requirements can be suitably  illustrated with two classical, useful examples, often
encountered an many practical situations: 
the mean value and the variance.

The empirical average $\overline{x}$ is a convergent, unbiased estimation of the mean value $\mu$ of its
underlying distribution: $\hat{\mu}=\overline{x}$. This statement can easily be demostrated, by evaluating the expectation value and variance of $\overline{x}$:
\begin{eqnarray}
E\left[ \overline{x} \right] & = & \frac{1}{N} \sum_{i=1}^N E\left[ x \right] \ = \ \mu \ , \label{eq:samplemean}
\\
V\left[ \overline{x} \right] & = & E\left[ (\overline{x}-\mu)^2 \right] \ = \ \frac{\sigma^2}{N} \ .
\end{eqnarray} 
In contrast, the empirical RMS of a sample is a biased, asymptotically unbiased estimator of the variance $\sigma^2$: this can be demonstrated
by first rewriting its square (also called sample variance) in terms of the mean value:
\begin{eqnarray}
{\rm RMS}^2 \ = \ \frac{1}{N} \sum_{i=1}^N \left(x_i-\overline{x}\right)^2 
              \ = \ \frac{1}{N}\sum_{i=1}^N\left(x_i-\mu\right)^2 - \left( \overline{x}-\mu\right)^2 \ ,
\end{eqnarray}
and so its expectation value is 
\begin{eqnarray}
E\left[{\rm RMS}^2\right] \ = \ \sigma^2 - V\left[\overline{x}\right] \ = \ \frac{N-1}{N}\sigma^2 \ ,
\end{eqnarray}
which, while properly converging to the true variance $\sigma^2$  in the $N\to\infty$ limit, 
systematically underestimates its value for a finite-sized sample.
One can instead define a modified estimator
\begin{eqnarray}
\frac{N}{N-1}{\rm RMS}^2 \ = \ \frac{1}{N-1} \sum_{i=1}^N \left(x_i-\overline{x}\right)^2 \ , \label{eq:samplerms}
\end{eqnarray}
that ensures, for finite-sized samples, an unbiased estimation of the variance.

In summary, consistent and unbiased estimators for the mean value $\mu$ and variance $\sigma^2$ of an unknown underlying
PDF can be extracted from a finite-sized sample realized out of this PDF:
\begin{eqnarray}
\hat{\mu} & = &  \frac{1}{N}\sum_{i=1}^{N} x_i \ , \label{eq:muhat}
\\
\hat{\sigma}^2 & = & \frac{1}{N-1}\sum_{i=1}^{N}\left(x_i-\hat{\mu}\right)^2 \ . \label{eq:sigmahat}
\end{eqnarray} 
As a side note, the $1/N$ and $1/(N-1)$ factors for $\hat{\mu}$ in Eq.~(\ref{eq:muhat}) and for $\hat{\sigma}^2$ in 
Eq.~(\ref{eq:sigmahat}) can
be intuitively understood as follows: while the empirical average be estimated even on the smallest sample consisting
of a single event,
at least two events are needed to estimate their empirical dispersion.

\subsubsection{Covariance, correlations, propagation of uncertainties}

These two classical examples discussed in~\ref{sec:musigma} refer to a single random variable.
In presence of several random variables, expressed as a random $n$-dimensional vector $\vec{x}=\left\{x_1,\dotsc ,x_n\right\}$, 
the discussion 
leads to the definition of the empirical covariance matrix, whose elements $\hat{C}_{ab}$ can be estimated on a sample
of $N$ events as
\begin{eqnarray}
\hat{C}_{ab} \ = \ \frac{1}{N-1} \sum_{i=1}^N \left(x_{a,i}-\hat{\mu_a}\right)\left(x_{b,i}-\hat{\mu_b}\right) \ .
\end{eqnarray}
(the $a,b$ indices run over random variables, $1\leq a,b\leq n$).
Assuming the covariance is known (i.e. by means of the estimator above, or from first principles), the variance 
of an arbitrary function of these random variables $f(\vec{x})$
can be evaluated from a Taylor-expansion around the mean values $\hat{\vec{\mu}}$ as
\begin{eqnarray}
f\left(\vec{x}\right) \ = \ 
f\left(\hat{\vec{\mu}}\right) + \sum_{a=1}^n \left.\frac{df}{dx_a}\right|_{\vec{x}=\hat{\vec{\mu}}} \left(x_a-\hat{\mu}_a\right) \ ,
\end{eqnarray}
which leads to $E\left[f(\vec{x})\right] \ \simeq \  f(\hat{\vec{\mu}})$; similarly,
\begin{eqnarray}
E\left[f^2(\vec{x})\right] \ \simeq \ 
f^2(\hat{\vec{\mu}}) + \sum_{a,b=1}^n\left.\frac{df}{dx_a}\frac{df}{dx_b}\right|_{\vec{x}=\hat{\vec{\mu}}} \hat{C}_{ab} \ ,
\end{eqnarray}
and thus the variance of $f$ can be estimated as
\begin{eqnarray} \label{eq:errorpropagation}
\hat{\sigma}^2_f \ \simeq \ \sum_{a,b=1}^n\left.\frac{df}{dx_a}\frac{df}{dx_b}\right|_{\vec{x}=\hat{\vec{\mu}}} \hat{C}_{ab} \ .
\end{eqnarray}
This expression in Eq.~(\ref{eq:errorpropagation}), called the error propagation formula,
allows to estimate the variance of a generic function $f(\vec{x})$
from the estimators of mean values and covariances.

A few particular examples of error propagation deserve being mentioning explicitly:
\begin{itemize}
	\item 	If all random variables $\left\{x_a\right\}$ are uncorrelated, the covariance matrix is diagonal,
		$C_{ab}=\sigma_a^2\delta_{ab}$ and the covariance of $f(\vec{x})$ reduces to
\begin{eqnarray}
\hat{\sigma}^2_f \ \simeq \ \sum_{a=1}^n\left.\left(\frac{df}{dx_a}\right)^2\right|_{\vec{x}=\hat{\vec{\mu}}}\hat{\sigma}_a^2  \ .
\end{eqnarray}
	\item	For the sum of two random variables $S=x_1+x_2$, the variance is 
\begin{eqnarray}
\sigma_S^2 \ = \ \sigma_1^2+\sigma_2^2+2C_{12} \ = \ \sigma_1^2+\sigma_2^2+2\sigma_1\sigma_2\rho_{12}, \label{eq:sumquadrature}
\end{eqnarray}
		and the corresponding generalization to more than
		two variables is straightforward: 
\begin{eqnarray}
\sigma_S^2 \ = \ \sum_{a,b} \sigma_a\sigma_b\rho_{ab} \ .
\end{eqnarray}
	In absence of correlations, one says that absolute 
		errors are added in quadrature: hence the expression 
		in Eq.~(\ref{eq:sumquadrature})  is often written as  $\sigma_S \ = \ \sigma_1 \oplus \sigma_2$.
	\item	For the product of two random variables $P=x_1x_2$, the variance is	
\begin{eqnarray}\label{eq:prodquadrature}
\left(\frac{\sigma_P}{P}\right)^2 \ = \ 
	\left(\frac{\sigma_1}{x_1}\right)^2 + 
	\left(\frac{\sigma_2}{x_2}\right)^2 + 
	2\frac{\sigma_1}{x_1}\frac{\sigma_2}{x_2}\rho_{12} \ ,
\end{eqnarray}
with its generalization to more than two variables:
\begin{eqnarray}
 \left(\frac{\sigma_P}{P}\right)^2 \ = \ \sum_{a,b} \frac{\sigma_a}{x_a}\frac{\sigma_b}{x_b}\rho_{ab} \ .
\end{eqnarray}
		In absence of correlations, one  says that relative errors are added in quadrature, and 
		Eq.~(\ref{eq:prodquadrature}) $\sigma_P/P = \sigma_1/x_1\oplus\sigma_2/x_2$.
	\item 	For a generic power law function, $Z=x_1^{n_1}x_2^{n_2}\dotsc$, if all variables are uncorrelated,
the variance is
\begin{eqnarray}
\frac{\sigma_Z}{Z} \ = \ n_1\frac{\sigma_1}{x_1}\oplus n_2\frac{\sigma_2}{x_2}\oplus\dotsc 
\end{eqnarray}
\end{itemize}

\section{A survey of selected distributions}

In this Section, a  brief description of distributions, often encountered in practical applications, is presented.
The rationale leading to this choice of PDFs  is driven 
either by their specific mathematical properties, and/or
in view of their common usage
in the modellingi of important physical processes; such features are correspondingly emphasized in the discussion.

\subsection{Two examples of discrete distributions: Binomial and Poisson}\label{sec:binomialPoisson}
\subsubsection{The binomial distribution}
Consider a scenario with two possible outcomes: ``success'' or ``failure'', with a fixed probability $p$ of
``success'' being realized (this is also called a Bernouilli trial). If $n$ trials are 
performed, $0\leq k\leq n$ may actually result in ``success'';
it is assumed that the sequence of trials is irrelevant, and only 
the number of ``success'' $k$ is considered of interest. 
The integer number $k$ follows the so-called Binomial distribution $P(k;n,p)$:
\begin{eqnarray}
P_{\rm binomial}\left(k;n,p\right) \ = \ \frac{n!}{k!(n-k)!}p^k\left(1-p\right)^{n-k} \ ,
\end{eqnarray}
where $k$ is the random variable, while $n$ and $p$ are parameters. The mean value  and variance are
\begin{eqnarray}
E\left[k\right] & = & \sum_{k=1}^n kP\left(k;n,p\right) \ = \ np \ , \nonumber
\\
V\left[k\right] & = & np(1-p) \ .
\end{eqnarray}
\begin{figure}[ht]
\begin{center}
\includegraphics[width=7.5cm]{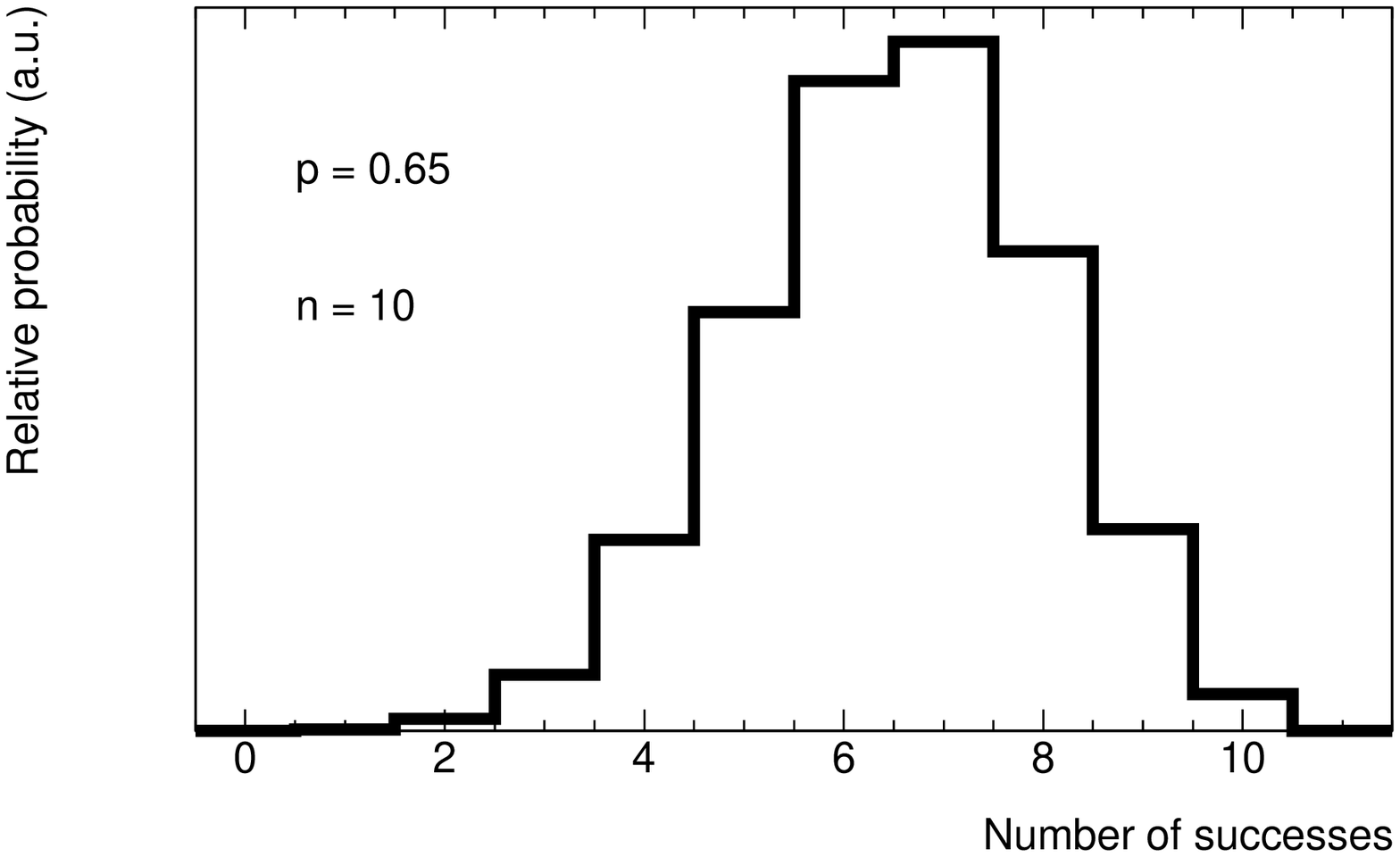}
\includegraphics[width=7.5cm]{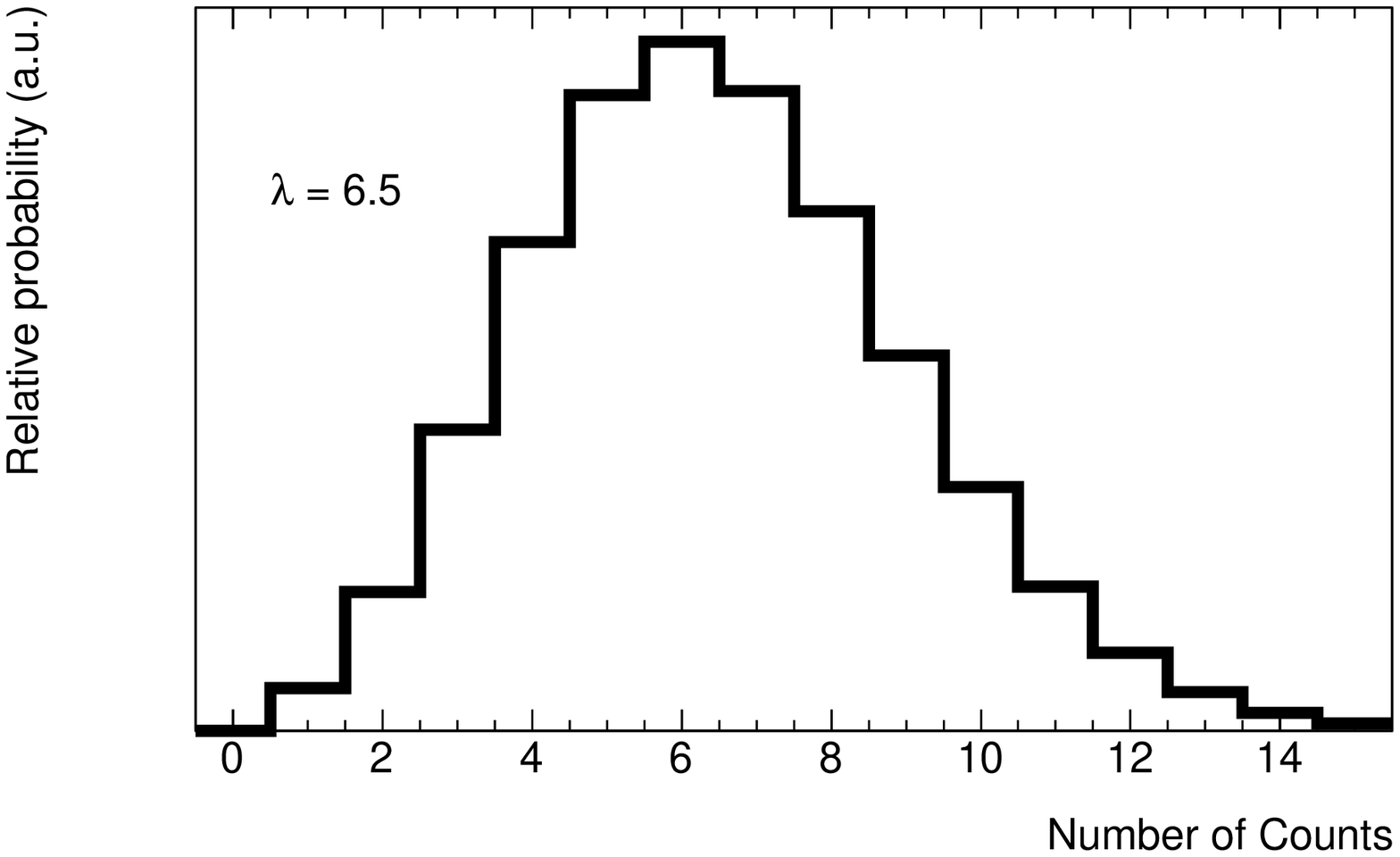}
\caption{Examples of a Binomial (left) and a Poisson (right) distributions, with
parameters $n=10$, $p=0.6$ and $\lambda=6.5$, respectively.}
\label{fig:binomialPoisson}
\end{center}
\end{figure}
\subsubsection{The Poisson distribution}
In the $n\to\infty$, $p\to 0$ limit (with $\lambda=np$ finite and non-zero) for the Binomial distribution,
the random variable $k$ follows the Poisson distribution $P(k;\lambda)$,
\begin{eqnarray}
P_{\rm Poisson}\left(k;\lambda\right) \ = \ \frac{\lambda^ke^{-\lambda}}{k!} \ ,
\end{eqnarray}
for which $\lambda$ is the unique parameter. For Poisson, the mean value and variance are the
same:
\begin{eqnarray}
E\left[k\right] \ = \ V\left[k\right] \ = \ \lambda \ .
\end{eqnarray}
The Poisson distribution, sometimes called law of rare events (in view of the $p\to 0$ limit), 
is a useful model for describing event-counting rates.
Examples  of a Binomial and a  Poisson distribution, for $n=10$, $p=0.6$ and for $\lambda=6.5$, respectively,
are shown on Figure~\ref{fig:binomialPoisson}. 

\subsection{Common examples of real-valued distributions}\label{sec:examples}
\begin{figure}[ht]
\begin{center}
\includegraphics[width=7.5cm]{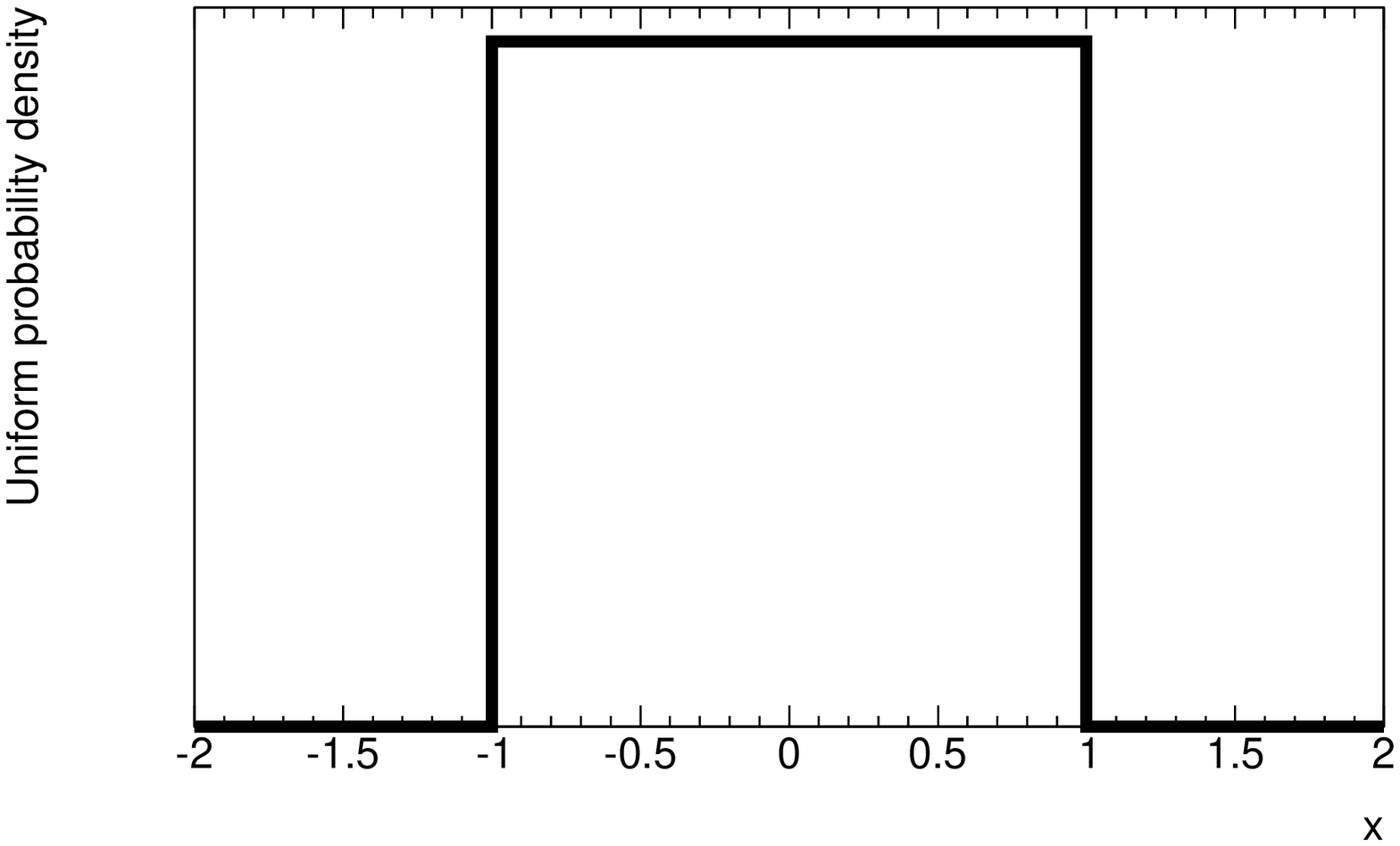}
\includegraphics[width=7.5cm]{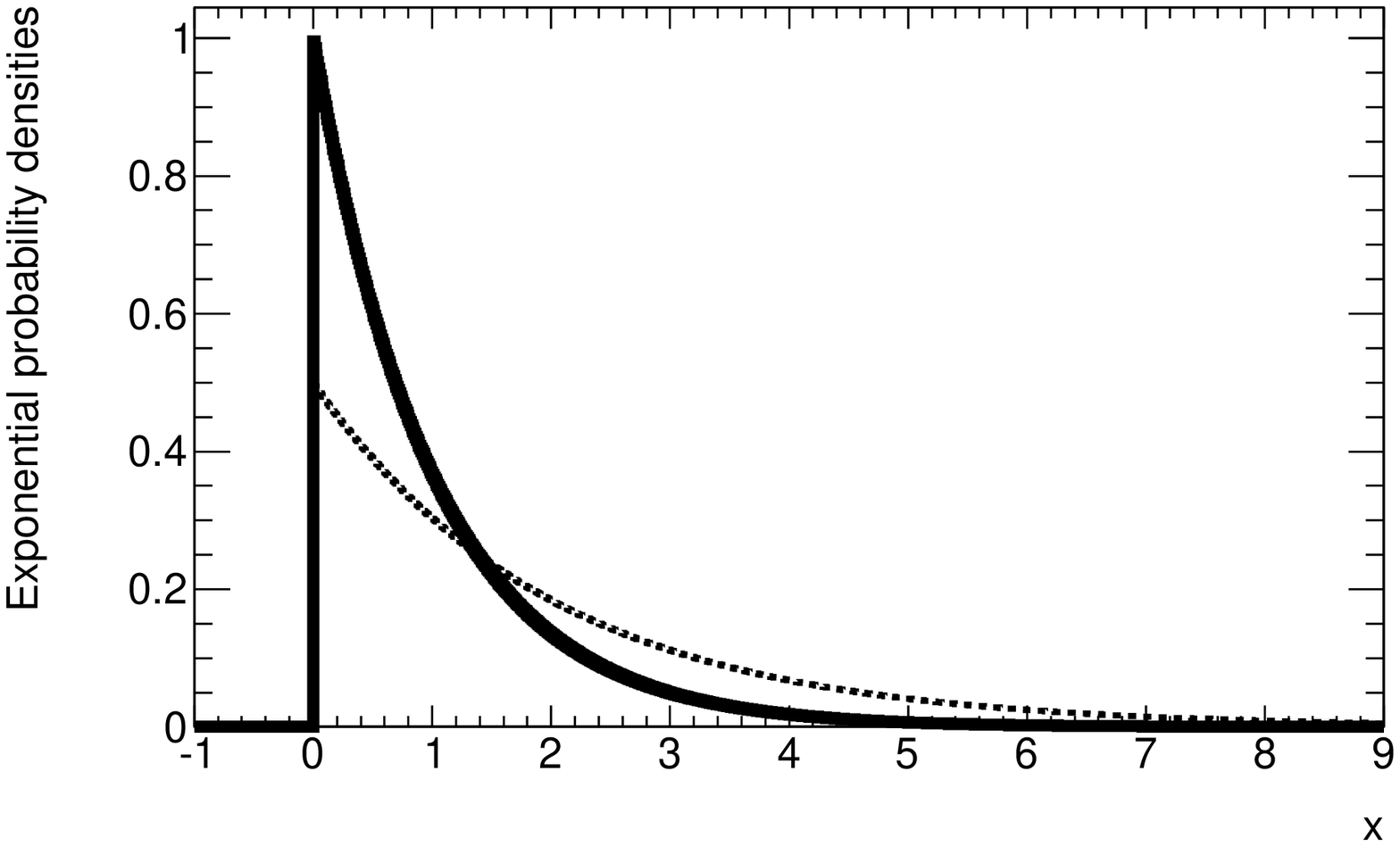}
\caption{Examples of a uniform (left) and an exponentially-decreasing (right) distribution. For the uniform distribution,
values used for the boundary parameters are $a=-1$, $b=1$; for the exponential, two values $\xi=1$ (solid line) and 
$\xi=2$ (dashed line) are used.}
\label{fig:uniformexponential}
\end{center}
\end{figure}

The first two continuous random variables discussed in this paragraph are the
uniform and the exponential distributions, illustrated in Figure~\ref{fig:uniformexponential}.
\subsubsection{The uniform distribution}
Consider a continuous random variable $x$, with a probability density $P(x;a,b)$ that is non-zero only inside
a finite interval $\left[a,b\right]$:
\begin{eqnarray}
P_{\rm uniform}\left(x;a,b\right) \ = \ 
\left\{ 
	\begin{array}{lcl}
		\frac{1}{b-a} & , & a\leq x\leq b \ ,
		\\
		0             & , & {\rm otherwise} \ .
	\end{array}
\right. 
\end{eqnarray}
For this uniform distribution, the mean value and variance are
\begin{eqnarray}
E\left[x\right] & = & \frac{a+b}{2} \ ,
\\
V\left[x\right] & = & \frac{(b-a)^2}{12} \ .
\end{eqnarray}
While not being the most efficient one, a straightforward Monte-Carlo generation approach 
would be based on a uniform distribution in the $\left[0,p\right]$ range,
and use randomly generated values of $x$ in the $\left[0,1\right]$ range as implementation of an accept-reject algorithm
with success probability $p$.
\subsubsection{The exponential distribution}
Consider a continuous variable $x$,
with a probability density $P(x;\xi)$ given by
\begin{eqnarray}
P_{\rm exponential}\left(x;\xi\right) \ = \ 
\left\{ 
	\begin{array}{lcl}
		\frac{1}{\xi}e^{-x/\xi} & , & x\geq 0 \ ,
		\\
		0             & , & {\rm otherwise} \ ,
	\end{array}
\right. 
\end{eqnarray}
whose mean value and variance are
\begin{eqnarray}
E\left[x\right] & = & \xi \ ,
\\
V\left[x\right] & = & \xi^2 \ .
\end{eqnarray}
A common application of this exponential distribution is the description of phenomena 
occuring independently at a constant rate, such as decay lengths and lifetimes.  
In view of the self-similar feature of the exponential function:
\begin{eqnarray}
P(t-t_0|t>t_0) \ = \ P(t) \ ,
\end{eqnarray}
the exponential distribution is sometimes said to be memoryless.

\subsubsection{The Gaussian distribution}
\begin{figure}[ht]
\begin{center}
\includegraphics[width=9.5cm]{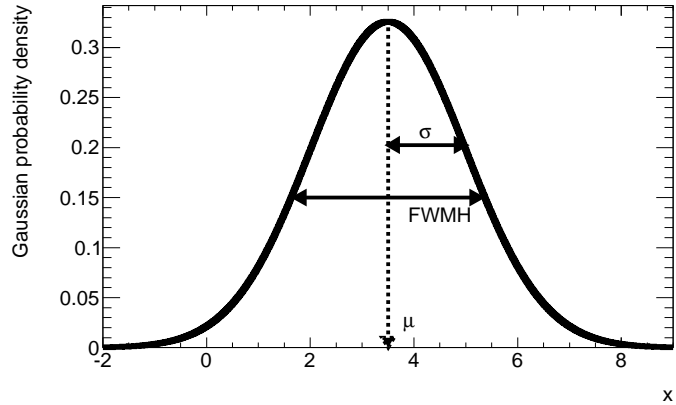}
\caption{A Gaussian PDF, with  parameter values $\mu=3.5$ and $\sigma=1.5$.}
\label{fig:gaussian}
\end{center}
\end{figure}
Now turn to the Normal (or Gaussian) distribution. Consider a random variable $x$, with probability density
\begin{eqnarray}
P_{\rm Gauss}\left(x;\mu,\sigma\right) \ = \ \frac{1}{\sqrt{2\pi\sigma}}e^{-\frac{(x-\mu)^2}{2\sigma^2}} \ ,
\end{eqnarray}
and with mean value and variance given by
\begin{eqnarray}
E\left[x\right] \ & = \ \mu \ ,
\\
V\left[x\right] \ & = \ \sigma \ .
\end{eqnarray}
The PDF corresponding to the special $\mu=0$, $\sigma=1$ case is usually
called ``reduced normal''.  

On purpose, the same symbols $\mu$ and $\sigma$ have been used both for the parameters of the Gaussian PDF and for 
the mean value and variance. This is an important feature: the Gaussian distribution is uniquely 
characterized by its first and second moments. For all Gaussians, the $\left[\mu-\sigma;\mu+\sigma\right]$
covers  $68.3\%$ of PDF integral, and is customary called a ``one-sigma interval''; similarly
for the two-sigma interval  and its $95.4\%$.

The dispersion of a peaked distribution is sometimes characterised in terms of its
FWHM (full width at half-maximum); for Gaussian distributions, this quantity is uniquely related  to its variance,
as ${\rm FWHM}=2\sqrt{2\ln{2}}\simeq 2.35\sigma$; Figure~\ref{fig:gaussian} provides a graphical illustration of the Gaussian PDF
and its parameters.

\begin{figure}[ht]
\begin{center}
\includegraphics[width=14.5cm]{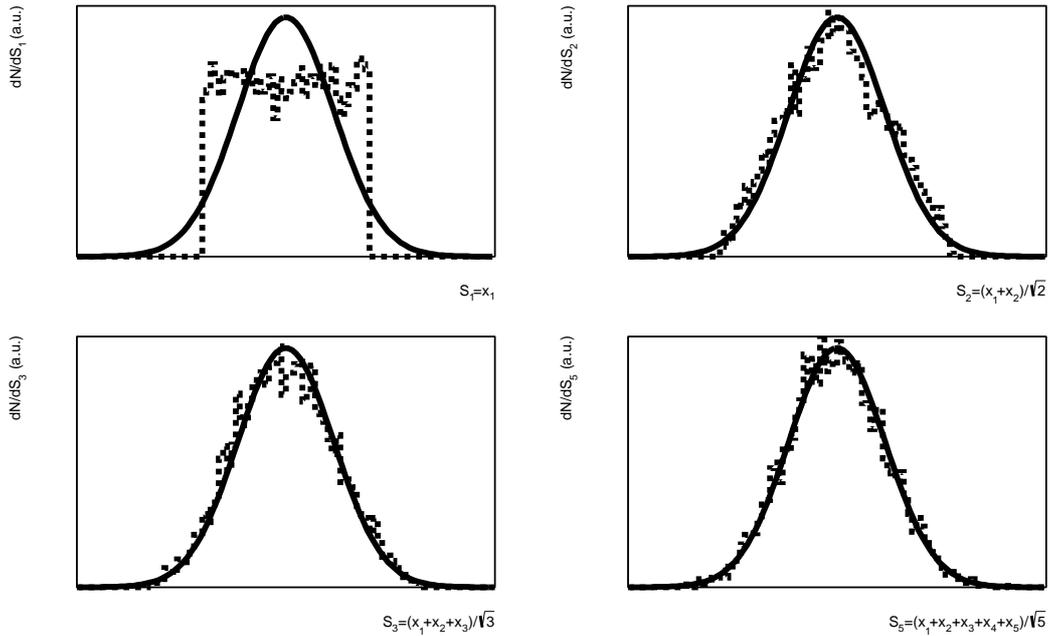}
\caption{A graphical illustration of the central limit theorem.
The top left plot compares the histogram (dashed curve) of a sample realization
from a uniform variable $x_1$, with a normal PDF (solid line) of same mean and variance; similarly, the plots on the top right, 
bottom left and bottom right plot the
corresponding histograms for 
the sums $S_2$, $S_3$ and $S_5$ of two, three and five reduced uniform variables $x_1,\dotsc ,x_5$, respectively. The
sequence of variables follow distributions that quickly converge to a reduced Gaussian.}
\label{fig:theoremCanvas}
\end{center}
\end{figure}

In terms of conceptual relevance and practical applications, the Gaussian certainly outnumbers
all other common distributions; this feature is largely due to the central limit theorem, which asserts 
that Gaussian distributions are the limit of
processes arising from multiple random fluctuations. Consider $n$ independent random variables
$\vec{x}=\left\{x_1,x_2,\dotsc ,x_n\right\}$, each with mean and variances $\mu_i$ and $\sigma_i^2$; the variable $S(\vec{x})$, 
built as the sum of reduced variables
\begin{eqnarray}
S \ = \ \frac{1}{\sqrt{n}}\sum_{i=1}^n\frac{x_i-\mu_i}{\sigma_i} \ ,
\end{eqnarray}
can be shown to have a distribution that, in the large-$n$ limit, converges to a reduced normal distribution,
as illustrated in Figure~\ref{fig:theoremCanvas} for the sum of up to five uniform distributions.
Not surprisingly, any measurement subject to multiple sources of fluctuations is 
likely to follow a distribution that can be approximated with a Gaussian distribution to a 
good approximation, regardless of the specific details of the processes at play. 

The Gaussian is also encountered as the limiting distribution for the Binomial and and Poisson ones, in the large $n$ and large
$\lambda$ limits, respectively:
\begin{eqnarray}
P_{\rm binomial}\left(k;n\to\infty,p\right) & \rightarrow &  P_{\rm Gauss}\left(k;np,np(1-p)\right) \ ,
\\
P_{\rm Poisson}\left(k;\lambda\to\infty\right) & \rightarrow & P_{\rm Gauss}\left(k;\lambda,\sqrt{\lambda}\right) \ .
\end{eqnarray}
Note that, when using a Gaussian as approximation, an appropriate continuity correction needs to be
taken into account: the range of the Gaussian extends to negative values, while Binomial and Poisson 
are only defined in the positive range.

\subsubsection{The $\chi^2$ distributions}
\begin{figure}[ht]
\begin{center}
\includegraphics[width=9.5cm]{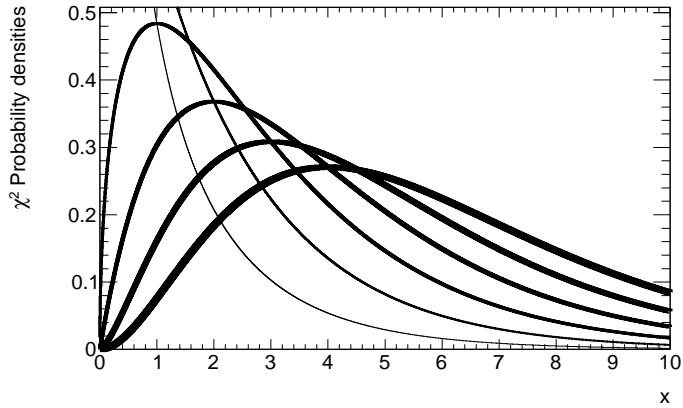}
\caption{The first six $\chi^2$ probability density functions, for integer numbers of degrees of freedom $n$.
The width of solid lines increases monotonically with
$n$ in the $n=1,\dotsc ,6$ range.} 
\label{fig:chi2Canvas}
\end{center}
\end{figure}

Now the $\chi^2$ (or chi-squared) distributions are considered. The following  PDF
\begin{eqnarray} \label{eq:chi2}
P_{\rm \chi^2}\left(x;n\right) \ = \ 
\left\{ 
	\begin{array}{lcl}
		\frac{x^{n/2-1}e^{-x/2}}{2^{n/2-1}\Gamma\left(\frac{n}{2}\right)} & , & x\geq 0 \ ,
		\\
		0             & , & {\rm otherwise} \ ,
	\end{array}
\right. 
\end{eqnarray}
with a single parameter $n$, and where $\Gamma\left(n/2\right)$ denotes the Gamma function,
has mean value and variance given by
\begin{eqnarray}
E\left[x\right] \ & = \ n \ , \nonumber
\\
V\left[x\right] \ & = \ 2n \ .
\end{eqnarray}
The shape of the $\chi^2$ distribution depends thus on $n$, as shown in Figure~\ref{fig:chi2Canvas} where the $\chi^2$ PDF, 
for the first six integer values
of the $n$ parameter, are shown.
It can be shown that the $\chi^2$ distribution
can be written as the sum of 
squares of $n$ normal-reduced variables $x_i$, each with mean $\mu_i$
and variance $\sigma_i^2$: 
\begin{eqnarray}
P_{\rm \chi^2}\left(x;n\right) \ = \ \sum_{i=1}^n \left(\frac{x_i-\mu_i}{\sigma_i}\right)^2 \ .
\end{eqnarray}
In view of this important feature of the $\chi^2$ distribution, the quantity $n$ is called
``number of degrees of freedom''; this name refers to the expected behaviour of a least-square fit,
where $n_d$ data points are used to estimate $n_p$ parameters; the corresponding number of degrees of freedom is $n_d-n_p$.
For a well-behaved fit,  the $\chi^2$  value should
follow a $\chi^2$ distribution. As discussed in~\ref{sec:chi2test}, the comparison of an observed $\chi^2$ 
value with its expectation, is an example  of goodness-of-fit test.

\subsubsection{The Breit-Wigner and Voigtian distributions}
The survey closes with the discussion of two  physics-motivated PDFs. The
first is the Breit-Wigner function, which
is defined as
\begin{eqnarray}
P_{\rm BW}\left(x;\Gamma,x_0\right) \ = \ \frac{1}{\pi} \frac{\Gamma/2}{\left(x-x_0\right)^2+\Gamma^2/4} \ ,
\end{eqnarray}
whose parameters are the  most probable value $x_0$ (which specifies the peak of the distribution),
and the FWHM $\Gamma$.
The Breit-Wigner is also called Lorentzian by physicists, and in matematics it is often referred to as the Cauchy distribution.
It has a peculiar feature, as a consequence of its long-range tails: the empirical average and empirical RMS are ill-defined
(their variance increase with the size of the samples), and cannot be used as estimators of the Breit-Wigner parameters.
The truncated mean and  interquartile range, which are obtained 
by removing events in the low and high ends of the sample, are  safer estimators of the Breit-Wigner parameters.

In the context of relativistic kinematics, the Breit-Wigner
function provides a good description of a resonant process
(for example the invariant mass of decay products from a resonant intermediate state); for a resonance,
the parameters $x_0$ and $\Gamma$ are referred
to as its mass and its natural width, respectively.

\begin{figure}[ht]
\begin{center}
\includegraphics[width=6.8cm]{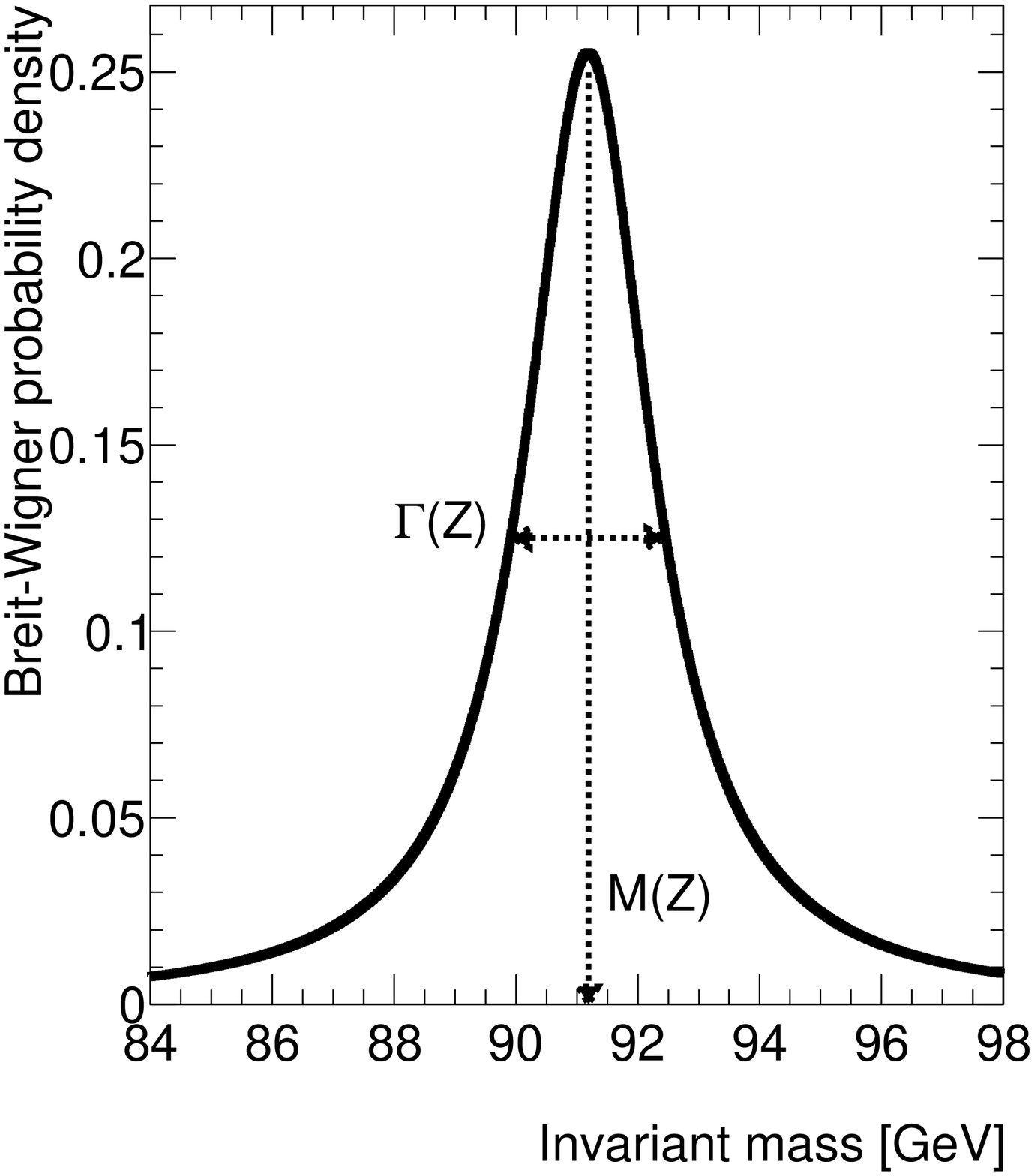}
\includegraphics[width=8.2cm]{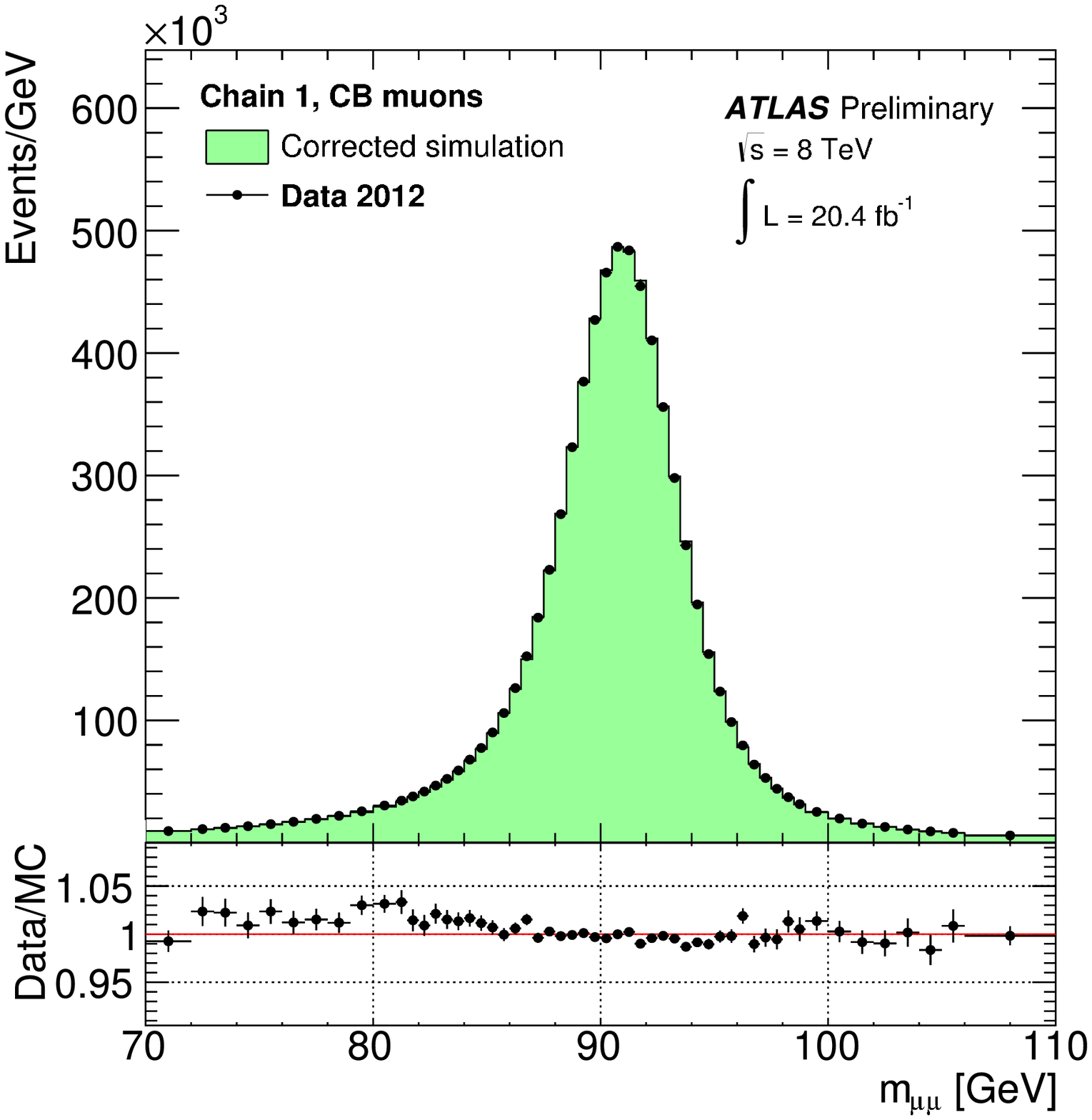}
\caption{Left: the probability density function for a Breit-Wigner distribution, using the PDG~\cite{bib:PDG} values for the
$Z$ boson mass and width as parameters. Right: the invariant dimuon mass spectrum
around the $Z$ boson mass peak; the figure is from an ATLAS measurement using 8 TeV data~\cite{bib:ATLASmumu}.} 
\label{fig:BWvoigt}
\end{center}
\end{figure}

Finally, the Voigtian function is the convolution of a Breit-Wigner with a Gaussian, 
\begin{eqnarray}
P_{\rm Voigt}\left(x;x_0,\Gamma,\sigma\right) \ = \ 
	\int_{-\infty}^{+\infty} dx^\prime P_{\rm Gauss}\left(x^\prime;0,\sigma\right)P_{\rm BW}\left(x-x^\prime;x_0,\Gamma\right) \ ,
\end{eqnarray}
and is thus a three-parameter distribution: mass $x_0$, natural width $\Gamma$ and resolution $\sigma$. While there is no
straightforward analytical form for the Voigtian, efficient numerical implementations are available, i.e.
the {\tt TMath::Voigt} member function in the ROOT~\cite{bib:ROOT} data analysis framework,
and the {\tt RooVoigtian} class in the {\tt RooFit}~\cite{bib:RooFit} toolkit for data modeling.
For values of $\Gamma$ and $\sigma$ sufficiently similar, the FWHM of a Voigtian can be approximated as a combination
of direct sums and sums in quadrature of the $\Gamma$ and $\sigma$ parameters. A simple, crude approximation yields :
\begin{eqnarray}
{\rm FWHM}_{\rm Voigt} \ \simeq \left[ \left(\Gamma/2\right) \oplus \sigma \right] + \Gamma/2  \ .
\end{eqnarray}
When the instrumental resolutions are sufficiently well described by a Gaussian, a Voigtian distribution 
 is a good model for
observed experimental distributions of a resonant process.
Figure~\ref{fig:BWvoigt} represents an analytical Breit-Wigner PDF 
(evaluated using the $Z$ boson mass and width as parameters), and a 
dimuon invariant mass spectrum around the $Z$ boson mass peak, as measured by the ATLAS experiment using 8 TeV 
data~\cite{bib:ATLASmumu}. 
The width of the observed peak 
is interpreted in terms of experimental resolution effects, as indicated by the good data/simulation agreement.
Of course, the ATLAS dimuon mass resolution is more complicated
than a simple Gaussian (as hinted by the slightly asymmetric shape of the distribution), therefore 
a simple Voigtian function would not have
reached the level of accuracy provided by the complete simulation of the detector resolution.

\section{The maximum likelihood theorem and its applications}\label{sec:MLtheorem}

The discussion in Section~\ref{sec:parameterestimation} was aimed at describing a few intuitive examples
of parameter estimation and their properties. Obviously, such case-by-case approaches are not general enough.
The maximum likelihood estimation (MLE) is an important general method for parameter
estimation, and is based  on properties following  the maximum likelihood (ML) theorem.

Consider a sample made of $N$ independent outcomes of random variables $\vec{x}$, arising from a $n$-parametric
PDF $P\left(\vec{x};\theta_1,\dotsc,\theta_n\right)$, and whose analytical dependence on 
variables and parameters is known, but for which the value(s) of at least one of its parameters $\theta$ is unknown.
With the MLE, these values are extracted from an analytical expression,
called the likelihood function, that has a functional dependence derived from the PDF,
and is designed to maximise the probability of realizing the observed outcomes.
The likelihood function ${\cal L}$ can be written as
\begin{eqnarray}\label{eq:likelihood}
{\cal L}\left(\theta_1,\dotsc,\theta_n;\vec{x}\right) \ = \ \prod_{i=1}^N P\left(\vec{x}_i;\theta_1,\dotsc,\theta_n\right) \ .
\end{eqnarray}
This notation shows implictly the functional dependence of the likelihood on the parameters $\theta$, and 
on the $N$ realizations $\vec{x}$ of the sample.
The ML theorem states that  the $\hat{\theta_1},\dotsc,\theta_n=\hat{\theta}_n$ values
that maximize ${\cal L}$, 
\begin{eqnarray}
{\cal L}\left(\hat{\theta}_1,\dotsc,\hat{\theta}_n\vec{x}\right) \ = \ 
	\left.{\rm max}\right|_{\theta}\left\{{\cal L}\left(\theta_1,\dotsc,\theta_n\right)\right\} \ . 
\end{eqnarray}
are estimators of the unknown 
parameters $\theta$, 
with variances $\hat{\sigma}_\theta$ that are extracted from the covariance of ${\cal L}$ around its maximum.

In a few cases, the MLE can be solved analytically.
A classical
example is the estimation of the mean value and variance of an arbitrary sample, that can be analytically derived under the 
assumption that the underlying PDF is a Gaussian. Most  often though, a  numerical study of
the likelihood around the $\theta$ parameter space is needed to localize the  $\hat{\theta}$ point that minimizes
$-\ln{{\cal L}}$ (the ``negative log-likelihood'', or NLL); this procedure is called a ML fit.

\subsection{Likelihood contours}\label{sec:contours}

Formally speaking, several conditions are required
for the ML theorem to hold. For instance, ${\cal L}$ has to be at least twice derivable
with respect to all its $\theta$ parameters; 
constraints on (asymptotical) unbiased behaviour and efficiency must be satisfied; and 
the shape of ${\cal L}$ around its maximum must be normally distributed. This last condition is particularly relevant,
as it ensures the accurate extraction of errors.
When it holds, the likelihood is said (in a slightly abusive manner) to have a ``parabolic'' shape
(more in reference to $-\ln{L}$ than to the likelihood itself), and its expansion
around $\hat{\theta}$ can be written as
\begin{eqnarray} \label{eq:parabolic}
f\left( \hat{\vec{\theta}} , \vec{\theta} , \mathbf{\Sigma} \right) \ = \ 
\frac{1}{\sqrt{2\pi}\left|\mathbf{\Sigma}\right|} 
\exp{\left\{-\frac{1}{2}\left( \hat{\vec{\theta}}_i - \vec{\theta}_i \right)\Sigma^{-1}_{ij}\left( \hat{\vec{\theta}}_j - \vec{\theta}_j \right)\right\}} \ .
\end{eqnarray} 
In Eq.~(\ref{eq:parabolic}) the covariance matrix $\mathbf{\Sigma}$  has been introduced, and its elements are given by :
\begin{eqnarray}
\mathbf{\Sigma}_{ij}^{-1} \ = \  -E\left[ \frac{\partial\ln{\cal L}}{\partial\theta_i} \frac{\partial\ln{\cal L}}{\partial\theta_j} \right] \ .
\end{eqnarray}
When moving away from  its maximum, the likelihood value decreases by amounts that depend on the covariance matrix elements:
\begin{eqnarray}
-2\Delta\ln{\cal L} \ = \ -2 \left[ \ln{\cal L}(\vec{\theta}) - \ln{\cal L}(\hat{\vec{\theta}}) \right] 
\ = \ \sum_{i,j}^{} \left( \theta_i - \hat{\theta}_i \right) \mathbf{\Sigma}_{ij}^{-1} \left( \theta_i - \hat{\theta}_j \right) \ .
\end{eqnarray}
This result, together with the result on error propagation in Eq.~(\ref{eq:errorpropagation}), indicates
that the covariance matrix defines contour 
maps around the ML point, corresponding to confidence intervals.
In the case of a single-parameter likelihood
${\cal L}(\theta)$, the interval contained in a
$-2\Delta\ln{\cal L}<1$ contour around $\hat{\theta}$ defines a $68\%$ confidence interval, 
corresponding to a $-\sigma_{\theta}\leq \theta-\hat{\theta}\leq \sigma_{\theta}$ range around the ML point;
in consequence the result of this MLE can be quoted as $\left(\hat{\theta}\pm\hat{\sigma}_\theta\right)$.

\subsection{Selected topics on maximum likelihood}

\subsubsection{Samples composed of multiple species}

In a typical scenario, the outcome of a random process may arise from multiple sources.
To be specific, consider that events in the data sample are a composition of two ``species'', called generically
``signal'' and ``background'' (the generalization to scenarios with more than two species is straighforward).
Each species is supposed to be realized from its own probability densities, yielding similar (but not identical) 
signatures in terms of the random variables in the data sample; it is precisely these residual differences in PDF shapes that
are used by the ML fit for a statistical separation of the sample into species. In the two-species example, the 
underlying total PDF is a combination
of both signal and background PDFs, and the corresponding likelihood function  is given by
\begin{eqnarray}\label{eq:MLfit1}
{\cal L}\left(\theta;\vec{x}\right) \ = \ \prod_{i=1}^N \left[ f_{\rm sig} P_{\rm sig}(\vec{x};\theta) + (1-f_{\rm sig}) P_{\rm bkg}(\vec{x};\theta) \right] \ ,
\end{eqnarray}
where $P_{\rm sig}$ and $P_{\rm bkg}$ are the PDFs for the signal and background species, respectively, and
the signal fraction $f_{\rm sig}$
is the parameter quantifying the signal purity in the sample: $0\leq f_{\rm sig}\leq 1$.  Note that, since  both
$P_{\rm sig}$ and $P_{\rm bkg}$ satisfy the PDF normalization condition from Eq.~(\ref{eq:normalization}), the 
total PDF used in Eq.~(\ref{eq:MLfit1})
is also normalized. It is worth mentioning that some of the parameters $\theta$ can be common to both signal and background
PDFs, and others may be specific to a single species.
Then, depending on the process
and the study under consideration, the signal fraction
can either have a known value, or belong to the set of unknown  parameters $\theta$ to be estimated in a ML fit. 

\subsubsection{ Extended ML fits}

In event-counting experiments, the actual number of observed events of a given species is a quantity of
interest; it is then convenient to treat the number of events as an additional parameter $\lambda$ of the likelihood function. 
In the case of a single species, this amounts to ``extending'' the likelihod,
\begin{eqnarray} \label{eq:EMLfit1}
{\cal L}\left(\lambda,\theta;\vec{x}\right) \ = \ \frac{\lambda^Ne^{-\lambda}}{N!}\prod_{i=1}^{N} P\left(\vec{x_i};\theta\right) \ .
\end{eqnarray}
where an additional multiplicative term, corresponding to the Poisson distribution (c.f. Section~\ref{sec:binomialPoisson}), has 
been introduced.
(the $N!$ term in the denominator can be safely dropped; a global factor has no impact on the
shape of the likelihood function nor on the ML fit  results). 
It is straightforward to verify that the Poisson likelihod in  Eq.~(\ref{eq:EMLfit1}) 
is maximal when $\hat{\lambda}=N$, as intended; now, if some of the PDFs also depend on $\lambda$, the value of
$\hat{\lambda}$ that maximises ${\cal L}$ may differ from $N$.
The generalization to more than one species is straightforward as well; for each species,
a multiplicative Poisson term is included in the extended likelihood, and the PDFs of each species are weighted by their
corresponding relative event fractions; in presence of two species, the extended version
of the likelihood in Eq.~\ref{eq:MLfit1} becomes:
\begin{eqnarray}\label{eq:EMLfit2}
{\cal L}\left(N_{\rm sig},N_{\rm bkg}\theta;\vec{x}\right) \ = \ (N_{\rm sig}+N_{\rm bkg})^Ne^{-(N_{\rm sig}+N_{\rm bkg})} 
		\prod_{i=1}^N \left[ N_{\rm sig} P_{\rm sig}(\vec{x};\theta) + N_{\rm bkg} P_{\rm bkg}(\vec{x};\theta) \right] \ .
\end{eqnarray}

\subsubsection{``Non-parabolic'' likelihoods,  likelihoods with multiple maxima}

As discussed in Section~\ref{sec:contours}, the condition in Eq.~(\ref{eq:parabolic}) about the asymptotically normal distribution 
of the likelihood around its maximum is crucial to ensure a proper interpretation of $-2\Delta\ln{\cal L}$ contours
in terms of confidence intervals. In this paragraph, two scenarios in which this condition can break down are discussed. 

\begin{figure}[ht]
\begin{center}
\includegraphics[width=7.5cm]{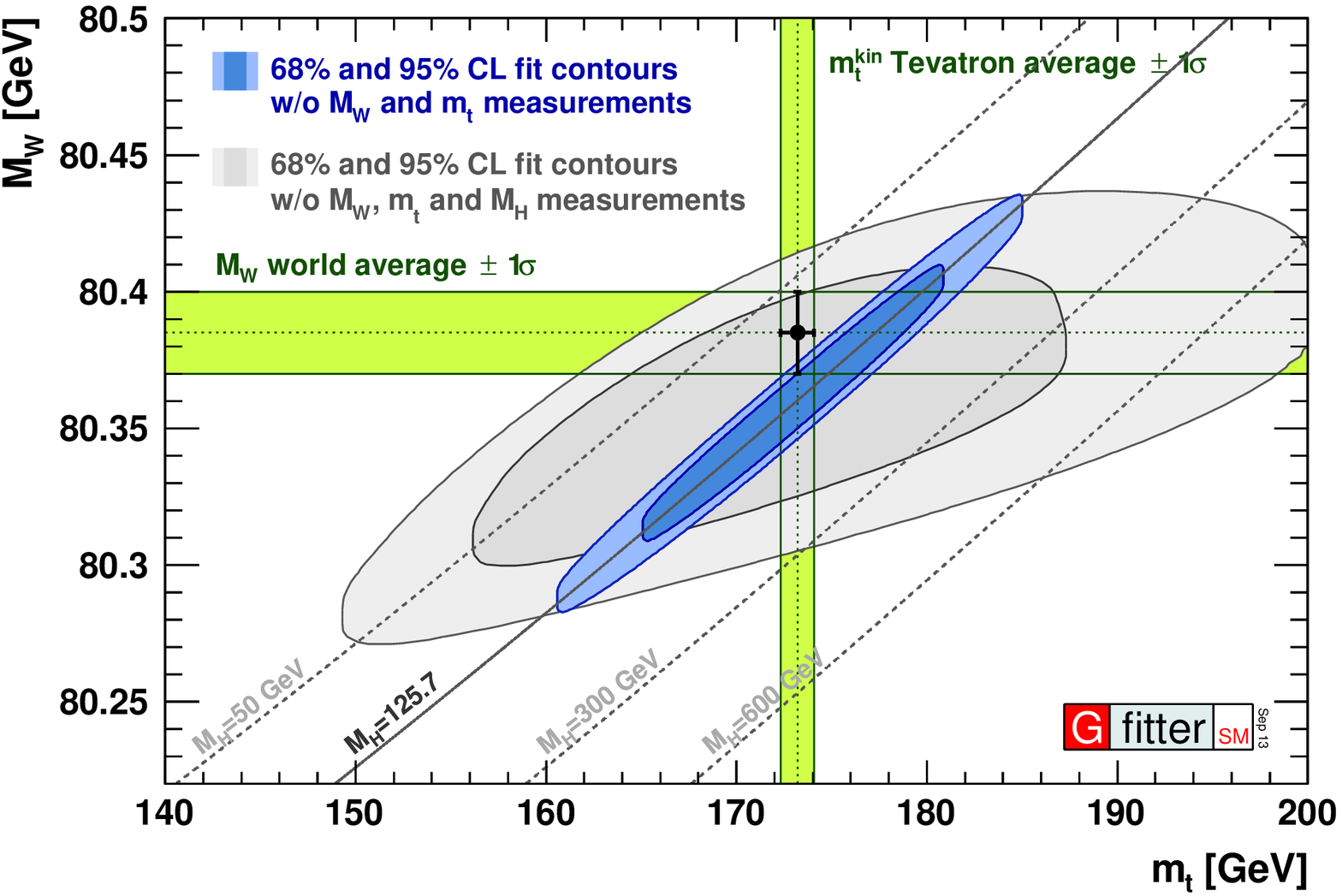}
\includegraphics[width=7.5cm]{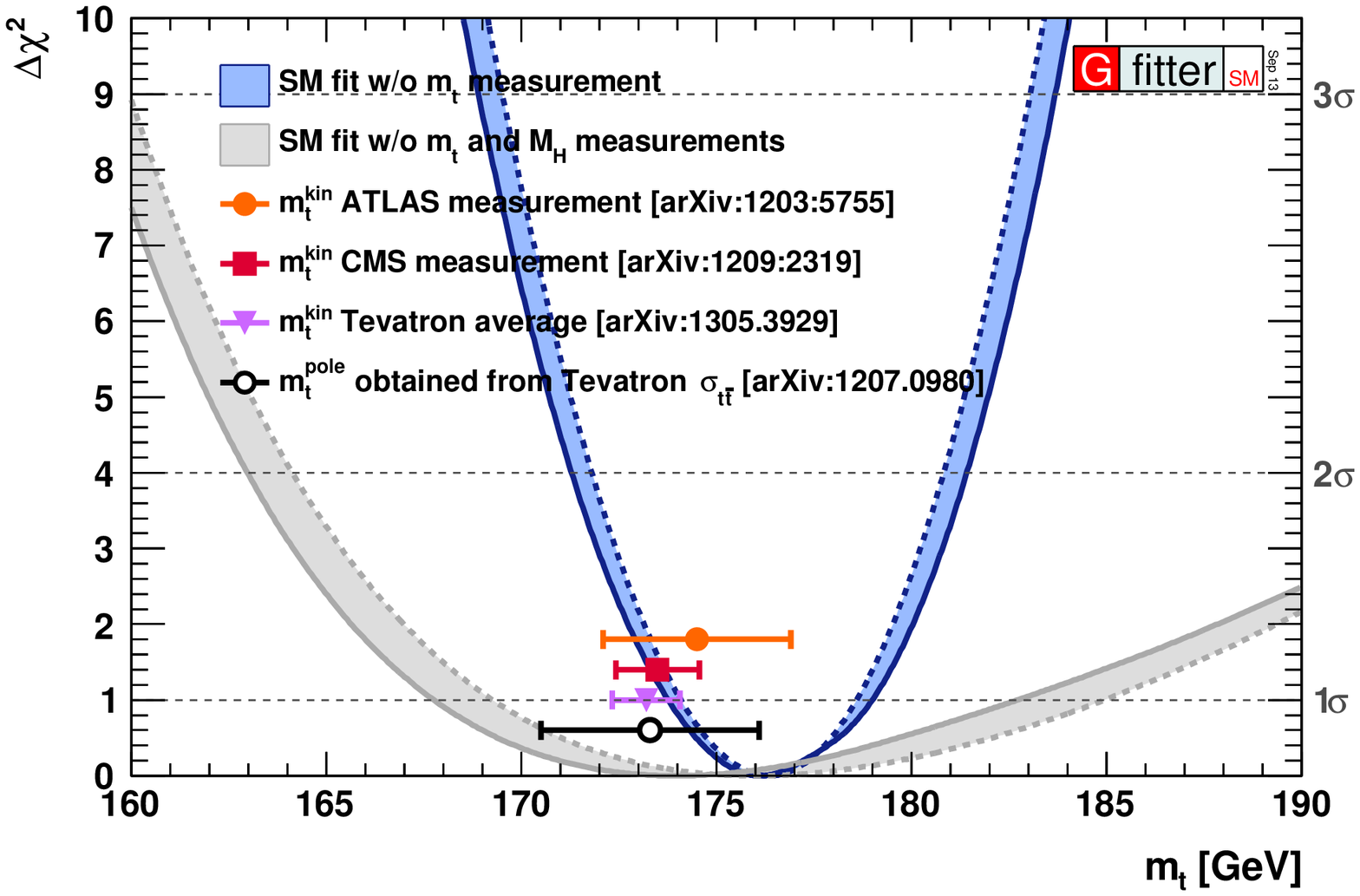}
\caption{ 
Left: two-dimensional likelihood contours
in the $m_w-m_t$ plane, obtained from a global electroweak fit (EW) performed by the
gFitter collaboration~\cite{bib:gFitter}. 
Center: the change in the EW likelihood as a function of $m_t$, expressed in terms of $\delta\chi^2=-2\Delta\ln{L}$.
Both plots illustrate the non-parabolic shape of the EW likelihood. 
}
\label{fig:gFitter}
\end{center}
\end{figure}

\begin{figure}[ht]
\begin{center}
\includegraphics[width=9.5cm]{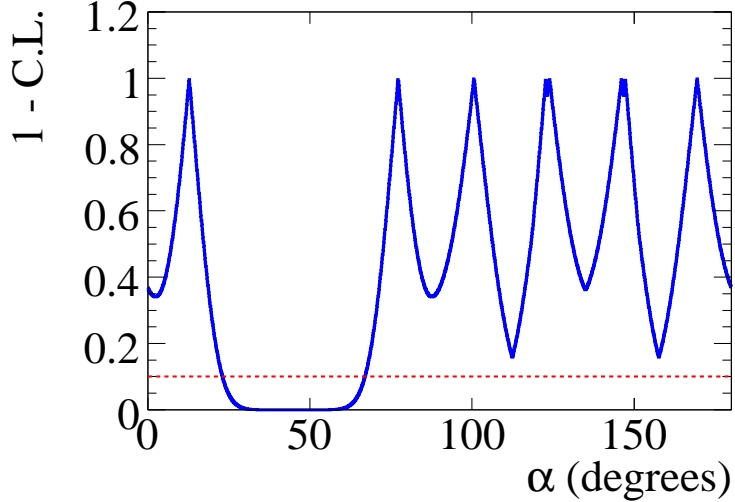}
\caption{ 
The change in likelihood, expressed in terms of a variant called ``confidence level'' CL, shown as a function of the CKM angle $\alpha$;
the definition of CL is such that ${\rm CL}=0$ at the solutions of the ML fit.
An eight-fold constraint on $\alpha$ is extracted from 
measurements in $B\rightarrow\pi\pi$ decays by the {\babar} experiment~\cite{bib:alphaBaBar}. 
The two
pairs of solutions around $\sim 130^\circ$ are very close in values, and barely distinguishable
in the figure. 
}
\label{fig:alphaBaBar}
\end{center}
\end{figure}

A first scenario concerns a likelihood that is not totally symmetric around its maximum. Such a feature may
occur, when studying  low-statistics data samples, in view of the Binomial or Poissonian 
behaviour of event-counting related quantities (c.f. Figure~\ref{fig:binomialPoisson}).
But it can also 
occur on larger-sized data samples, indicating that the model has a limited
sensitivity to the parameter $\theta$ being estimated, or as a consequence of strong
non-linear relations in the likelihood function. As illustration,  
examples  of two- and one-dimensional likelihood
contours,  with clear non-parabolic shapes, 
are shown in 
Figure~\ref{fig:gFitter}.

Also, the likelihood function may possess one or more local maxima, or even completely degenerate maxima. There are various
possible  sources
for such degeneracies. For example in models with various species, if some  PDF shapes are not different enough
to provide a robust statistical discrimination among their corresponding species. For example, if swapping the PDFs for
a pair of species yields
a sufficiently good ML fit, a local maximum may emerge;
on different sample realizations, the roles of local and global maxima may alternate among the correct and swapped 
combinations.
The degeneracies could also arise as a reflexion of explicit, physical symmetries in the model: for example,
time-dependent asymmetries in $B^0\rightarrow\pi^+\pi^-$ decays are sensitive to the 
CKM angle $\alpha$, but the physical observable is a function of $\cos{2(\alpha+\delta)}$, with an addtional phase 
$\delta$ at play; in consequence, the model brings up to eight indistinguishible solutions for the CKM angle $\alpha$, as
illustrated in Figure~\ref{fig:alphaBaBar}. 

In all such cases, the (possibly disjoint) $-2\Delta\ln{\cal L}<1$ interval(s)
around the $\hat{\theta}$ central value(s)  cannot be simply reported with a symmetric uncertainty $\hat{\sigma}_\theta$ only.
In presence of a single, asymmetric solution,  the measurement can be  quoted with 
asymmetric errors, i.e. $\hat{\theta}^{+\sigma_{+}}_{-\sigma_{-}}$, 
or better yet, by providing the detailed shape of $-2\Delta\ln{\cal L}$ as a function of the estimated parameter $\theta$.
For multiple solutions, more information needs to be provided:  for example, amplitude analyses (``Dalitz-plot'' analyses)
produced by the B-factories {\babar} and Belle, often reported the complete covariance matrices around each local solution
(see e.g.~\cite{bib:KspipiBaBar,bib:KspipiBelle} as examples).

\subsubsection{Estimating efficiencies with ML fits}

\begin{figure}[ht]
\begin{center}
\includegraphics[width=11.5cm]{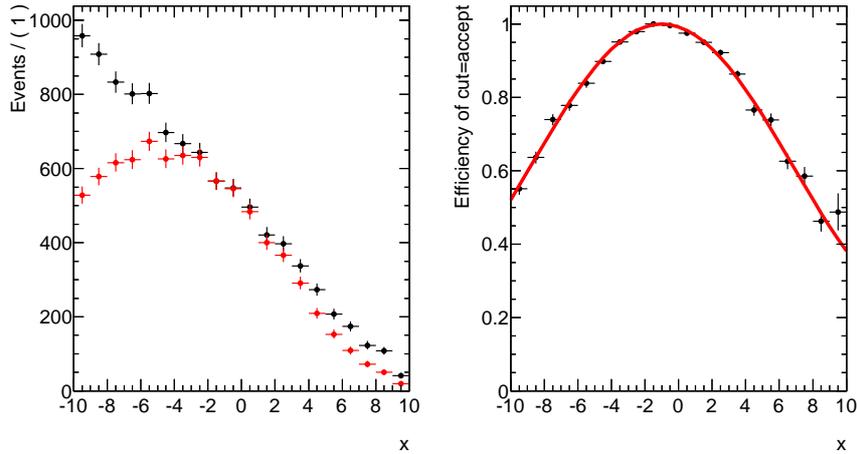}
\caption{ Left: the frequency of events in a sample, as a function  of a variable $x$. The sample contains two categories of 
events: ``accepted'' and ``rejected''. The red dots indicate the bins of the histogram 
for those ``accepted'' only, and the black dots for the two categories cumulated.
 Right:  the efficiency as a parametric function of $x$, with parameter values
extracted from a ML fit to the data sample. Figure extracted from the {\tt RooFit} user's guide~\cite{bib:RooFit}.}
\label{fig:efficiencyCanvas}
\end{center}
\end{figure}

Consider a process with two possible outcomes: ``yes'' and ``no''. The intuitive estimator of the efficiency $\varepsilon$ is a 
simple ratio, expressed  in terms of the number of outcomes $n_{\rm yes}$  and $n_{\rm no}$ of each kind:
\begin{eqnarray}
\hat{\varepsilon} \ = \ \frac{n_{\rm yes}}{n_{\rm yes}+n_{\rm no}} \ ,
\end{eqnarray}
for which the variance is given by
\begin{eqnarray}
V\left[\hat{\varepsilon}\right] \ =  \ \frac{\hat{\varepsilon}(1-\hat{\varepsilon})}{n} \ ,
\end{eqnarray}
where $n=n_{\rm yes}+n_{\rm no}$ is the total number of outcomes realized.
This estimator $\hat{\varepsilon}$ clearly breaks down for low $n$, and in the very low or very high efficiency regimes.
The MLE technique offers a robust approach to estimate efficiencies:  consider a PDF $P(x;\theta)$ to model the sample, and 
include in it  an additional discrete, bivariate random variable 
$c=\left\{{\rm yes}, {\rm no}\right\}$, so that the PDF becomes
\begin{eqnarray}
P\left(x,c;\theta\right) \ = \ \delta(c-{\rm yes})\varepsilon(x,\theta) + \delta(c-{\rm no})\left[1-\varepsilon(x,\theta)\right] \ .
\end{eqnarray}
In this way, the efficiency is no longer a single number, but a function of $x$ (plus some parameters $\theta$ that may 
be needed to
characterize its shape). With this function, the efficiency can be extracted in different $x$ domains, as illustated
in Figure~\ref{fig:efficiencyCanvas}, or can be used to produce a multidimensional efficiency map.

\subsection{Systematic uncertainties}

As discussed in Section~\ref{sec:contours}, in MLE the covariance of ${\cal L}$ is the estimator of statistical uncertainties.
Other potential sources of (systematical) uncertainties  are usually at play, and need to be quantified.
For this discussion, a slight change in the notation with respect to Eq.(\ref{eq:likelihood}) is useful; in this 
notation, the likelihood function 
is now written as
\begin{eqnarray}
{\cal L}\left( \mu_1,\dotsc,\mu_p \ , \ \theta_1,\dotsc,\theta_k ; \vec{x}\right) \ , 
\end{eqnarray}
where the parameters are explicitly partitioned into a set 
$\mu_1,\dotsc,\mu_p$, called parameters of interest (POI), that correspond to the
actual quantities that are to be estimated; and a set $\theta_1,\dotsc,\theta_k$, called
nuisance parameters (NP), that represent potential sources of systematic biases: if
inaccurate or wrong values are assigned to some NPs, the shapes of the PDFs can be distorted, the estimators
of POIs can become biased.
The systematic uncertainties due to NPs are usually classified in two categories. 

The first category refers to
``Type-I'' errors, for which the sample (or other control data samples) can (in principle) provide information on the NPs
under consideration, and are (in principle) supposed to decrease with sample size. 
The second category refers to
``Type-II'' errors, which arise from incorrect assumptions in the model (i.e. a choice of inadequate functional
dependences in the PDFs), or uncontrolled features in data
that can not be described by the model, like the presence of unaccounted species. Clearly, for Type-II errors 
the task of assigning systematic uncertainties to them may not be well-defined, and may spoil
the statistical interpretation of errors in termos of confidence levels.

\subsubsection{The profile-likelihod method}

To deal with Type-I  nuisance parameters, a popular approach is to use the so-called profile-likelihood method.
This approach consists of  assigning a specific likelihood to the nuisance parameters, so that  the original likelihood
is modified to have two different components:
\begin{eqnarray}\label{eq:profileLikelihood}
{\cal L}\left(\mu,\theta\right) \ = \ {\cal L}_\mu\left(\mu,\theta\right){\cal L}_\theta\left(\theta\right) \ . 
\end{eqnarray}
Then, for a fixed value of $\mu$, the likelihood is maximized with respect to the nuisance $\theta$; the
sequencial outcome of this procedure, called profile likelihood, is a function that depends only on $\mu$: it is then 
said that the nuisance
parameter has been profiled out of the likelihood. 

As an example, consider 
the measurement of the cross-section of a generic process, $\sigma\left({\rm initial}\to{\rm final}\right)$. If only
a fraction of the processes is actually detected, the
efficiency $\varepsilon$ of reconstructing the final state is needed to convert the
observed event rate $\hat{N}_{\rm event}$ into a measurement $\hat{\sigma}$. This efficiency is clearly a nuisance:
a  wrong value of $\varepsilon$ directly
affects the value of $\hat{\sigma}$, regardless of how accurately $\hat{N}_{\rm event}$ may have been measured.
By estimating $\hat{\varepsilon}$  on a quality control sample (for example, high-statistics simulation,
or a high-purity control data sample), the impact of this
nuisance can be attenuated. For example, an elegant analysis would  produce a simultaneous fit to the data and control samples, so that
the values and uncertainties of NPs are estimated in the ML fit, and are correctly propagated to the
values and variances of the POIs.

As another example, consider the search for a resonance (``a bump'') over
a uniform background. If the signal fraction is very small, the width $\Gamma$ of the bump cannot 
be directly estimated on data, and the value used in the signal PDF has to be inferred from external sources. 
This width is clearly a
nuisance: using an overestimated value would translate into an underestimation of the signal-to-background ratio,
and thus an increase in the variance of the signal POIs, and possibly biases in their central values as well,
i.e. the signal rate would tend to be overestimated. Similar considerations can be applied in case of underestimation
of the width. 
If an estimation 
$\hat{\Gamma}\pm\hat{\sigma}_\Gamma$ of the width is available, this information can be implemented as in
Eq.~(\ref{eq:profileLikelihood}), by using a Gaussian
PDF, with mean value $\hat{\Gamma}$ and width $\hat{\Gamma}$, in the ${\cal L}_\Gamma$ component of the likelihood. This term
acts as a penalty in the ML fit, and thus constraints the impact of the nuisance $\Gamma$ on the POIs.

\section{Multivariate techniques}

Often, there are large regions in sample space where backgrounds are overwhelming, and/or signals are absent.
By restricting the data sample
to ``signal-enriched'' subsets of the complete space, the loss of information may be minimal, and other advantages may
compensate the potential losses: in particular for multi-dimensional samples, it can be difficult to characterize the 
shapes in regions away from the core, where the event
densities are low; also reducing the sample size can relieve speed and memory consumption in numerical
computations.

The simplest method of sample reduction is by requiring a set of variables to be restricted into finite
intervals. In  practice, such ``cut-based'' selections appear at many levels in the definition of sample space:
thresholds on online trigger decisions,   filters at various levels of data acquisition,
removal of data failing quality criteria... But at more advanced stages of a data analysis, such ``accept-reject'' sharp selections
may have to be replaced by more sophisticated procedures, generically called multivariate techniques. 

A multi-dimensional ML fit is an example of a multivariate technique. For a MLE to be considered,
a key requirement is to ensure a robust knowledge of all PDFs over the space of random variables. 

Consider a set 
of $n$ random variables $\vec{x}=\left\{x_1,x_2\dotsc,x_n\right\}$.
If all variables are shown to be uncorrelated,
their  $n$-dimensional PDF is 
completely determined by the product of their $n$ one-dimensional PDFs; now,  if variables are
correlated, but their correlation patterns are completely linear, one can instead use variables $\vec{y}$,
linear combinations of $\vec{x}$ obtained by diagonalizing the inverse covariance.
For some  non-linear  correlation patterns, it may be possible to find
analytical descriptions; for instance, the (mildly) non-linear correlation
pattern represented in Figure~\ref{fig:pdf2DCanvas}, was produced with the {\tt RooFit}
package, by applying the 
{\tt Conditional} option in {\tt RooProdPdf} to build a product of PDFs. In practice, this
elegant approach cannot be easily extended to more than two dimensions, and is not guaranteed to reproduce
complex, non-linear patterns. In such scenarios, the approach of dimensional reduction can potentially  bring more effective
results.

A typical scenario for dimensional reduction is when several variables 
carry common information (and
thus exhibit strong correlations), together with some diluted (but relevant) pieces of independent information. An example 
is the characterization of
showers in calorimeters;  for detectors with good transverse and/or longitudinal segmentation,
the signals deposited in nearby calorimetric channels can be used to reconstruct details from the shower development;
for example, a function that combines informations from longitudinal and transverse shower shapes, 
can  be used to discriminate  
among electromagnetic and hadronic showers.

The simplest algorithm for dimensional reduction is the Fisher discriminant: it is a
linear function of variables, with coefficients
adjusted to match  an optimal criterion, called separation among two species, which is
is the ratio of the variance between the species to the variance within the species, and can be expressed in a close,
simple analytical form. 

\begin{figure}[ht]
\begin{center}
\includegraphics[width=9.5cm]{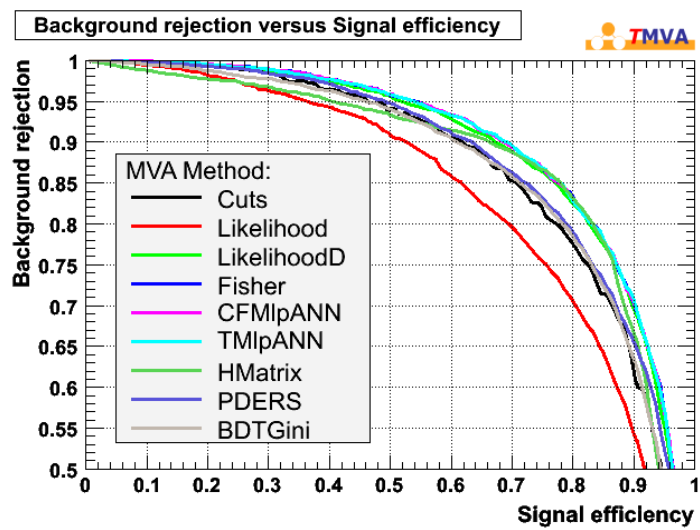}
\caption{ 
A figure-of-merit comparison between various different multivariate analyzer methods using the TMVA
package. The  training was performed on two simulated samples (``signal'' and ``background''), each consisting of four 
linearly correlated Gaussian-distributed variables.
The various lines indicate the trade-off between signal efficiency and background rejection.
The figure is taken from~\cite{bib:TMVA}.
}
\label{fig:TMVA}
\end{center}
\end{figure}

In presence of more complex, non-linear correlation patterns, a large variety of techniques and tools are available.
The TMVA~\cite{bib:TMVA} package is a popular implementation of dimensional-reduction algorithms;
other than linear and likelihood-based discriminants, it provides easy training and testing methods for
artificial neural networks and (boosted) decision trees, which 
are among those most often encountered in HEP analyses. As a general rule, a multivariate analyzer uses
a collection of variables, realized on two different samples (corresponding to ``signal'' and ``background'' species), to 
perform a method-dependent 
training, guided by some optimization criteria; then performances of the trained analyzer are evaluated  on independent realizations
of the species (this distinction between the training and testing stages is crucial to avoid ``over-training'' effects).
Figure~\ref{fig:TMVA} shows a graphical representation
of a figure-of-merit comparison of various analyzers implemented in TMVA.

\section{Statistical hypothesis testing}

Sections~\ref{sec:paramestimation} and~\ref{sec:MLtheorem} mainly discussed procedures for extracting numerical
information from data samples; that is, to perform measurements and report them in terms of
values and uncertanties.  The next step in data analysis
is to extract qualitative information from data: this is the domain of statistical hypothesis testing. 
The analytical tool to assess the agreement of an hypothesis with observations  is called a test statistic,
(or a statistic in short);
the outcome of a test is given in terms of a $p$-value, or probability of a ``worse''
agreement than the one actually  observed.

\subsection{The $\chi^2$ test}\label{sec:chi2test}

For a set of $n$ independent measurements $x_i$, their deviation with respect to predicted values $\mu_i$, expressed in 
units of their variances $\sigma_i$ is
called the $\chi^2$ test, and is defined as 
\begin{eqnarray}
\chi^2 \ = \ \sum_{\rm i=1}^{n}\left( \frac{x_i-\mu_i}{\sigma_i}\right)^2 \ .
\end{eqnarray}
An ensemble of $\chi^2$ tests is a random variable that, as mentioned in Section~\ref{sec:examples}, follows
the $P_{\chi^2}$ distribution (c.f. Eq.~\ref{eq:chi2}) for $n$ degrees of freedom. Its expectation value is $n$
and its variance $2n$; therefore
one does expect the $\chi^2$ value observed in a given test not to deviate much from 
the number of degrees of freedom, and so this value can be used to probe the agreement of prediction 
and observation. More precisely, one expects $68\%$ of tests to be contained within
a $n\pm\sqrt{2n}$ interval, and the $p$-value, or  probability of having a test with values larger than a
given $\chi^2$ value is
\begin{eqnarray}\label{eq:chi2test}
p \ = \ \int_{\chi^2}^{+\infty} dq \ P_{\chi^2}\left(q ; n\right) \ .
\end{eqnarray}
Roughly speaking, one would tend to be suspicious of small observed $p$-values, as they
may  indicate a trouble, either with the prediction or the data
quality. 
The interpretation of  the observed $p$-value (i.e. to decide whether it is too small or large enough)
is an important topic, and is discussed below in a more general approach.
 
\subsection{General properties of hypothesis  testing}\label{hypotesting}

Consider two mutually excluding hypotheses $H_0$ and $H_1$, that may describe some data sample; the hypothesis testing procedure
states how robust is $H_0$ to describe
the  observed data,
and how incompatible is $H_1$ with the observed data. 
The hypothesis $H_0$ being tested is called the ``null'' hypothesis, while
$H_1$ is the ``alternative'' hypothesis,

Note that in the context of the search for a (yet unknown) signal, the null $H_0$ corresponds to a ``background-only'' scenario,
and the alternative $H_1$ to ``signal-plus-background''; while in the context of excluding a (supposedly
inexistent) signal, the roles of the null and alternative hypotheses  are reversed: the null $H_0$ is ``signal-plus-background''
and the alternative $H_1$ is ``background-only''. 

The sequence of a generic test, aiming at accepting (or rejecting) the null hypothesis $H_0$ by
confronting it to a data sample, can be sketched as follows:
\begin{itemize}
	\item	build a test statistic $q$, that is, a function that reduces a data sample to a single numerical value;
	\item	define a confidence interval $W \ \rightarrow \ \left[q_{\rm lo}:q_{\rm hi}\right]$; 
	\item	measure $\hat{q}$;
	\item	if $\hat{q}$ is contained in $W$, declare the null hypothesis accepted; otherwise, declare it rejected. 
\end{itemize}
To characterize the outcome of this sequence, two criteria are defined:
a ``Type-I error''  is incurred in, if $H_0$ is rejected despite being true; 
while	a ``Type-II error'' is incurred in, if  $H_0$ is accepted  despite being false (and thus $H_1$ being true).
The rates of Type-I and Type-II errors are called $\alpha$ and 
$\beta$ respectively, and are determined  by integrating
the $H_0$ and $H_1$ probability densities over the confidence interal $W$,
\begin{eqnarray}
1-\alpha & = & \int_{W} dq{\cal P}\left(q|H_0\right) \ ,
\nonumber \\
\beta & = & \int_{W} dq{\cal P}\left(q|H_1\right) \ .
\end{eqnarray}
The rate $\alpha$ is also called ``size of the test'' (or size, in short), as  fixing $\alpha$ determines
the size of the confidence interval $W$. Similarly, $1-\beta$ is also called ``power''.
Together, size and power  characterize the performance of a test statistic; the Neyman-Pearson lemma
states that the optimal statistic is the likelihood ratio $q_\lambda$, 
\begin{eqnarray}
q_\lambda \left( {\rm data} \right) \ = \ \frac{{\cal L}\left({\rm data}|H_0\right)}{{\cal L}\left({\rm data}|H_1\right)} \ .
\end{eqnarray}
The significance of the test is given by the  $p$-value,
\begin{eqnarray}\label{eq:pvalue}
p \ = \ \int_{\hat{q}}^{+\infty} dq {\cal P}\left(q|H_0\right) \ . 
\end{eqnarray}
which is often quoted in terms of ``sigmas'', 
\begin{eqnarray}
p \ = \ \int_{n\sigma}^{+\infty} dz \frac{1}{\sqrt{2\pi}}e^{-z^2/2} \ = \ 1 - \frac{1}{2} {\rm erf}\left(\frac{n}{\sqrt{2}}\right) \ ,
\end{eqnarray}
so that for example a $p<0.0228$ outcome can be reported as   a ``two-sigma'' effect. Alternatively, it is common
practice to quote the complement of the $p$-value as a confidence level (C.L.).

The definition of a $p$-value as in Eq.~(\ref{eq:pvalue}) 
(or similarly in Eq.~(\ref{eq:chi2test}) for the example for a $\chi^2$ test)  
is clear and unambiguous.
But interpretation of $p$-values is partly subjective: the convenience of a numerical ``threshold of tolerance'' 
may depend on the kind of hypothesis being tested, or on common practice. In HEP usage, three different traditional
benchmarks are conventionally employed:
\begin{itemize}
	\item	in exclusion logic, a $95\%$ C.L. threshold on a signal-plus-background test to claim exclusion;
	\item	in discovery logic, a three-sigma threshold ($p<1.35\times 10^{-3}$)  on a background-only test
		to claim ``evidence'';
	\item	and a five-sigma threshold ($p<2.87\times 10^{-7}$)  on the background-only test
		is required to reach the ``observation'' benchmark.
\end{itemize}

\subsection{From LEP to LHC:  statistics in particle physics}
\begin{figure}[ht]
\begin{center}
\includegraphics[width=4.9cm]{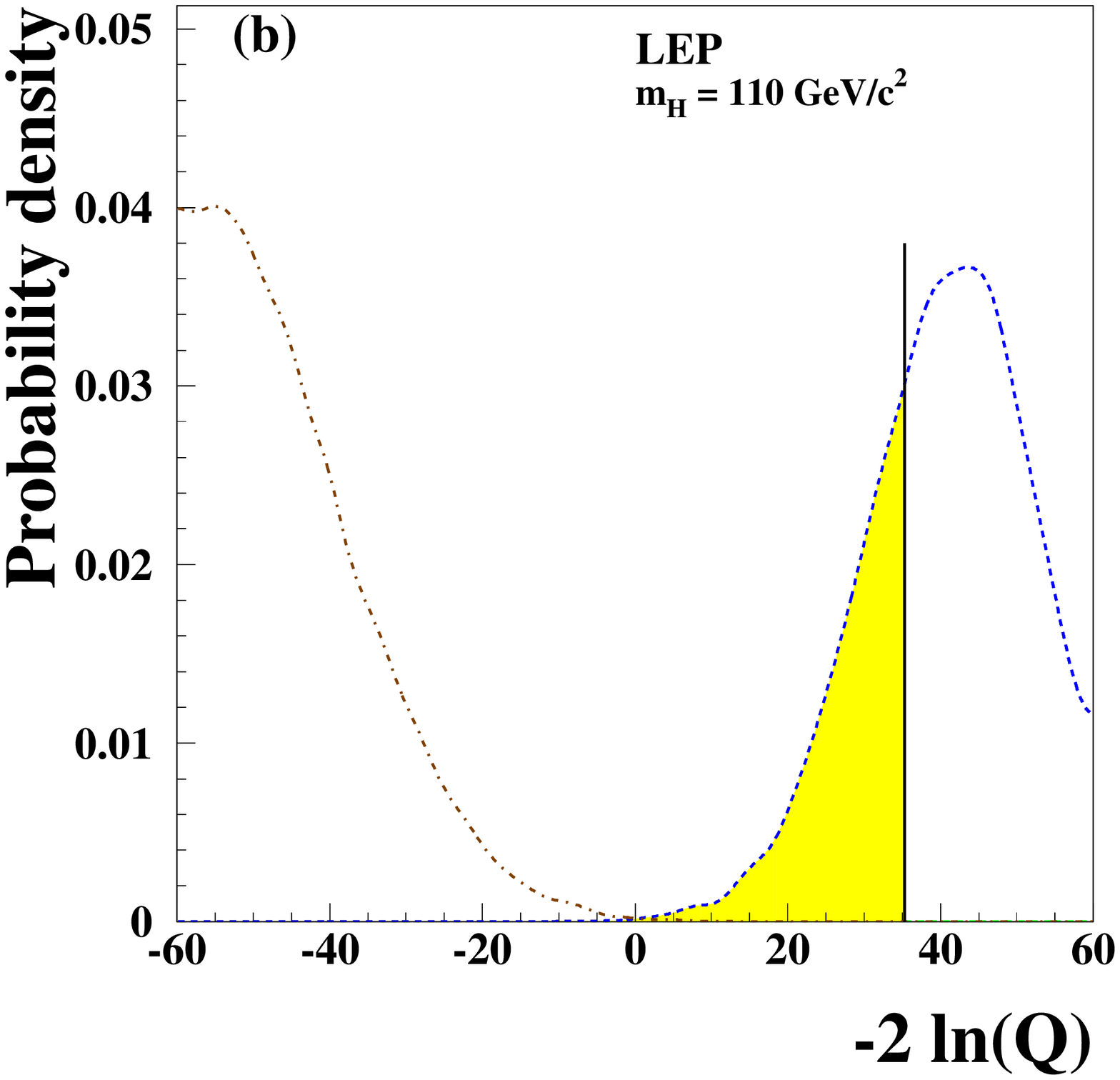}
\includegraphics[width=4.9cm]{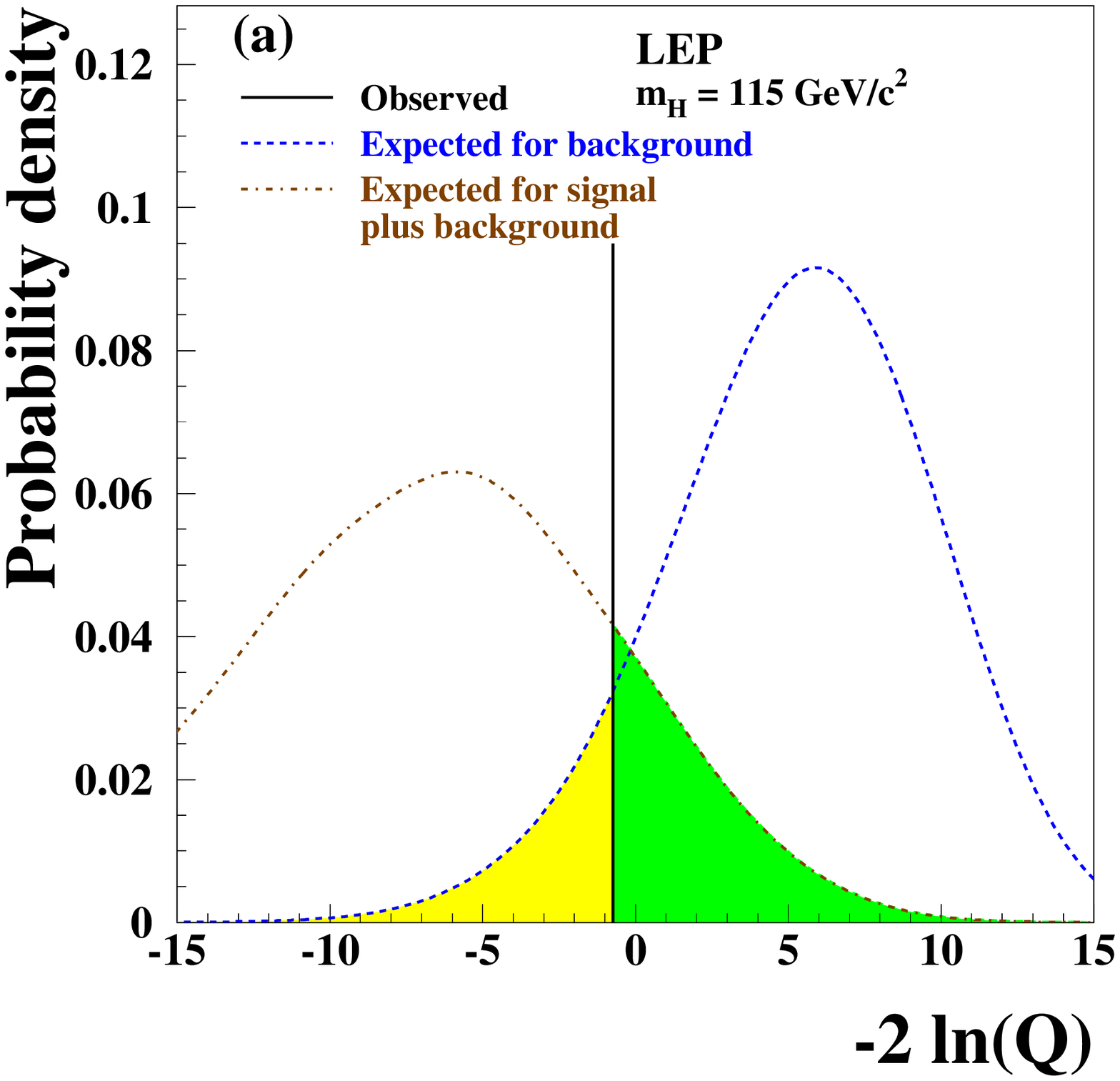}
\includegraphics[width=4.9cm]{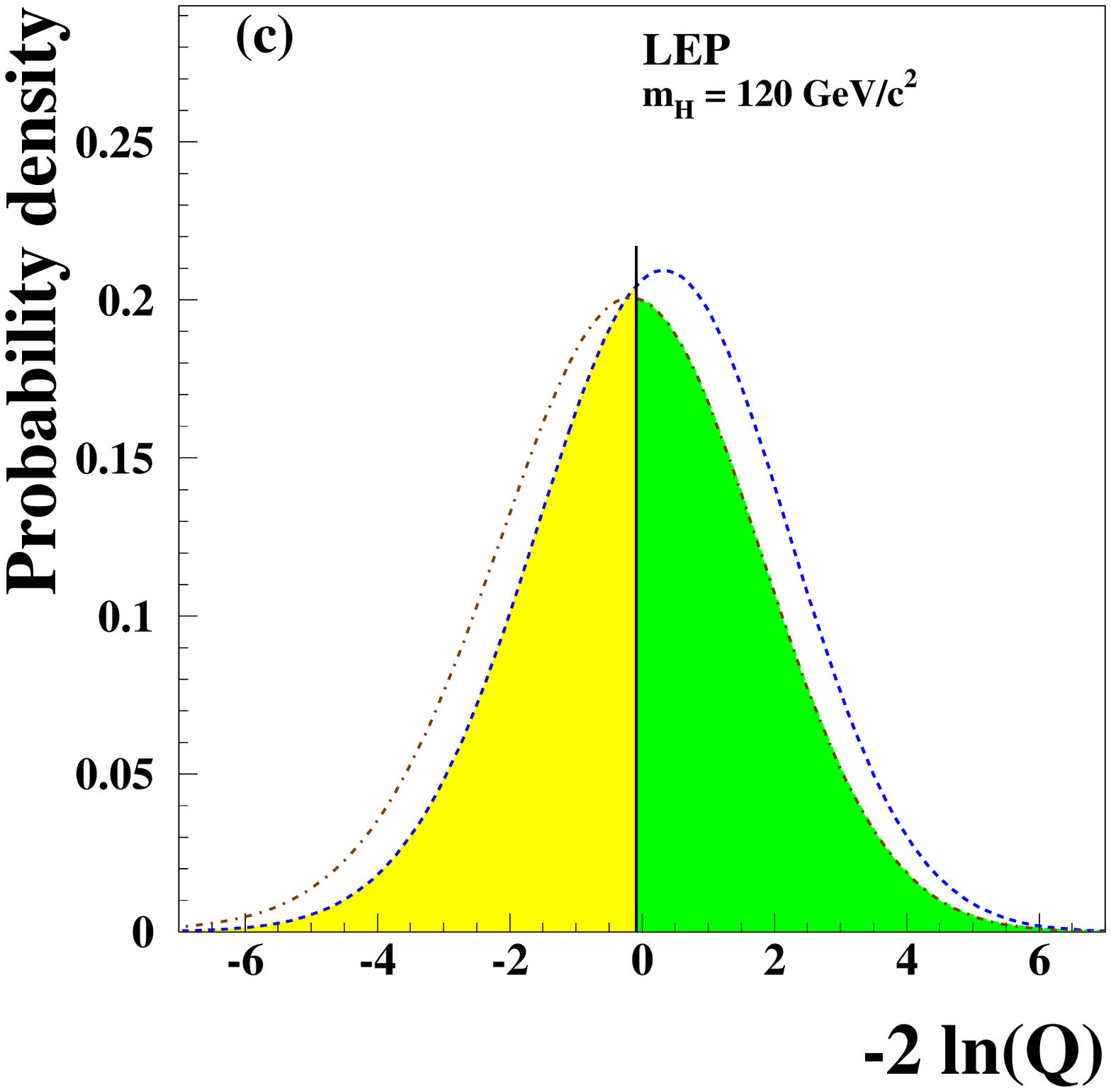}
\caption{ 
Combined results from the search for a SM Higgs boson, performed by the four LEP experiments.
From left to right,  the $m_H=110,115,120$ GeV hypotheses for the Higgs mass are used.
The dashed blue  and dashed red lines correspond to the PDFs for the background-only and signal-plus-background hypotheses,
respectively.
The observed values of the test statistic $-2\ln{Q}$ are marked by black vertical lines. 
The yellow areas indicate the $1-{\rm CL}(b)$ values for the background-only hypothesis,
and the green areas the  ${\rm CL}(s+b)$ value for signal-plus-background. The three figures are taken from~\cite{bib:SMHiggsLEP}.
}
\label{fig:LEP1D}
\end{center}
\end{figure}

\begin{figure}[ht]
\begin{center}
\includegraphics[width=9.5cm]{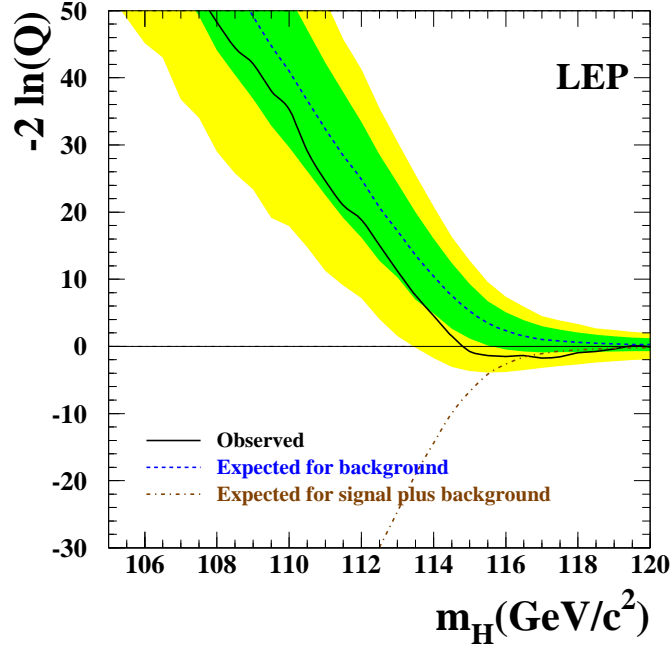}
\caption{ 
The test statistic $-2\ln{Q}$ as a function of the Higgs boson mass 
$m_H$, obtained by combining the data of the four LEP experiments. 
The dashed line is the mean value of the background-only distribution at each mass $m_H$, and the
green and yellow areas represent $68\%$ and $95\%$ contours for the background-only distribution at each mass $m_H$.
The black line follows the $-2\ln{Q}$ value observed in data as a function of $m_H$. Figure taken from~\cite{bib:SMHiggsLEP}.
}
\label{fig:LEP2D}
\end{center}
\end{figure}

In experimental HEP, there is a tradition of reaching consensual agreement on the  choices of test statistics. 
The goal is to ensure that, in the   combination of results from different samples and instruments,
the detector-related components (specific to each experiment) factor out from the physics-related observables (which
are supposed to be universal).
For example
in the context of searches for the Standard Model (SM) Higgs boson, 
the four LEP experiments agreed on analyzing
their data using the following template for their  likelihoods:
\begin{eqnarray}
{\cal L}\left(H_1\right) & = & 	\prod_{a=1}^{N_{\rm ch}} {\cal P}_{\rm Poisson}\left( n_a,s_a+b_a \right)
				\prod_{j=1}^{n_a} \frac{s_a{\cal S}_a(\vec{x_j})+b_a{\cal B}_a(\vec{x_j})}{s_a+b_a} \ ,
\nonumber \\
{\cal L}\left(H_0\right) & = & \prod_{a=1}^{N_{\rm ch}}{\cal P}_{\rm Poisson}\left( n_a,b_a \right)
				\prod_{j=1}^{n_a} {\cal B}_a(\vec{x_j}) \ .
\end{eqnarray} 
where $N_{\rm ch}$ is the number of Higgs decay channels studied, $n_a$ is the observed number of event candidates
in channel $a$, ${\cal S}_a$ and $s_a$ (${\cal B}_a$ and $b_a$) are the PDF and event yield for the signal (background) 
species in that channel.
Also, the test statistic $\lambda$, derived from a likelihood ratio, is 
\begin{eqnarray}
\lambda \ = \ -2\ln{Q} \ , \ {\rm with} \ Q & = & \frac{{\cal L}\left(H_1\right)}{{\cal L}\left(H_0\right)} \ ;
\end{eqnarray}
so that roughly speaking, positive values of $\lambda$ favour a ``background-like'' scenario, and negative ones are more in
tune with a  ``signal-plus-background'' scenario; values close to zero indicate poor sensitivity to distinguish
among the two scenarios.
The values use to test these two hypotheses are:
\begin{itemize}
	\item	under the background-only hypothesis, ${\rm CL}(b)$ is the probability of
		having a $-2\ln{Q}$ value larger than 
		the observed one;
	\item	under the signal+plus+background hypothesis, ${\rm CL}(s+b)$ is the probability of
		having a $-2\ln{Q}$ value larger than 
		the observed one. 
\end{itemize}
Figure~\ref{fig:LEP1D} shows, for three different Higgs mass hypotheses, 
the $-2\ln{Q}$ values from the combination of the four LEP experiments
in their searches for the SM Higgs boson, overlaid with the expected ${\rm CL}(s+b)$ and  $1-{\rm CL}(b)$
distributions.
Figure~\ref{fig:LEP2D} shows the evolution of $-2\ln{Q}$ values  as a function of the hypothetized Higgs boson mass;
(as stated in the captions, the color conventions in the one- and two-dimensional plots are different)
note that for masses below $\sim115$ GeV, a positive value of the $-2\ln{Q}$ test statistic would have provided evidence 
for a signal; and sensitivity is quickly lost above that mass.
\subsubsection{The modified ${\rm CL}(s)$ hypothesis testing}
The choice of ${\rm CL}(s+b)$ to test the signal-plus-background hypothesis, while suitably  defined
as a $p$-value.
may drag some subjective concern: in case of a fortuitous simultaneous downward fluctuation in both signal and background,
the standard $95\%$ benchmark may lead to an exclusion of the signal, even if the sensitivity is poor.


A modification of the exclusion benchmark called ${\rm CL}(s)$, has been introduced
in this spirit\cite{bib:Read}, and is defined
 as
\begin{eqnarray}\label{eq:CLs}
{\rm CL}(s) \ = \ \frac{ {\rm CL}(s+b) }{1- {\rm CL}(b)} \ .
\end{eqnarray} 
This test, while not corresponding to a $p$-value (a ratio of probabilities is not a probability), has the desired property of
protecting against downwards fluctuations of the background, and is commonly used in exclusion results,
including the searches for the SM Higgs boson from 
the Tevatron and LHC experiments.
\subsubsection{Profiled likelihood ratios}
Following the recommendations from the LHC Higgs Combination Group~\cite{bib:ATLASCMS}, the
ATLAS and CMS experiments have agreed on using a common test statistic, called profiled likelihood ratio and defined
as
\begin{eqnarray}
\tilde{q}_\mu\left(\mu\right) \ = \ -2\ln{\frac{{\cal L}\left(\mu,\hat{\hat{\theta}}\right)}{{\cal L}\left(\hat{\mu},\hat{\theta}\right)}}
\ , \ {\rm with} \ 0\leq\hat{\mu}\leq\mu \ ,
\end{eqnarray}
where the PIO $\mu=\sigma/\sigma_{\rm SM}$ is the ``signal strength modifier'', or Higgs signal rate expressed in 
units of the SM predicted rate, $\hat{\hat{\theta}}$ are the fitted values of the NPs at fixed values of the signal strength, and 
$\hat{\mu}$ and $\hat{\theta}$ are the fitted values when both $\mu$ and NPs are all free to vary in the ML 
fit~\footnote{The test statistic actually reported in ~\cite{bib:ATLASCMS} is slightly different than the one described here,
but in the context of these notes this subtlety can be disregarded.}.
The lower constraint on $0\leq\hat{\mu}\leq\mu$ ensures that the signal rate is positive, 
and the upper constraint imposes that an upward fluctuation would not disfavor 
the SM signal hypothesis.

For an observed statistic value $\hat{\tilde{q}}_\mu$, the  
$p$-values for testing the signal-plus-background and background-only hypotheses, $p(s+b)$ and $p(b)$, are
\begin{eqnarray}
p(s+b) & = & \int_{\hat{\tilde{q}}_\mu}^\infty dq P\left( q ; \mu=\hat{\mu}, \hat{\theta}\right) \ ,
\\
1-p(b) & = & \int_{\hat{\tilde{q}}_\mu}^\infty dq P\left( q ; \mu=0 , \hat{\hat{\theta}}\right) \ .
\end{eqnarray}
and the results on searches are reported both in terms of the exclusion significance using
the  ${\rm CL}(s)$ observed and expected values, and the observation significance expressed in
terms of the ``local''~\footnote{In short, the observation significance must be corrected for the trials 
factor, or ``look-elsewhere effect'';
in the case of the search for the SM Higgs boson in a wide Higgs mass interval, this effect translates into a 
decrease of the significance, as different Higgs masses are tested using independent data, and thus the 
probability of observing a signal-like
fluctuation depends on the mass interval studied and the experimental resolution.} $p(b)$ expected and observed
values.

\begin{figure}[ht]
\begin{center}
\includegraphics[width=7.5cm]{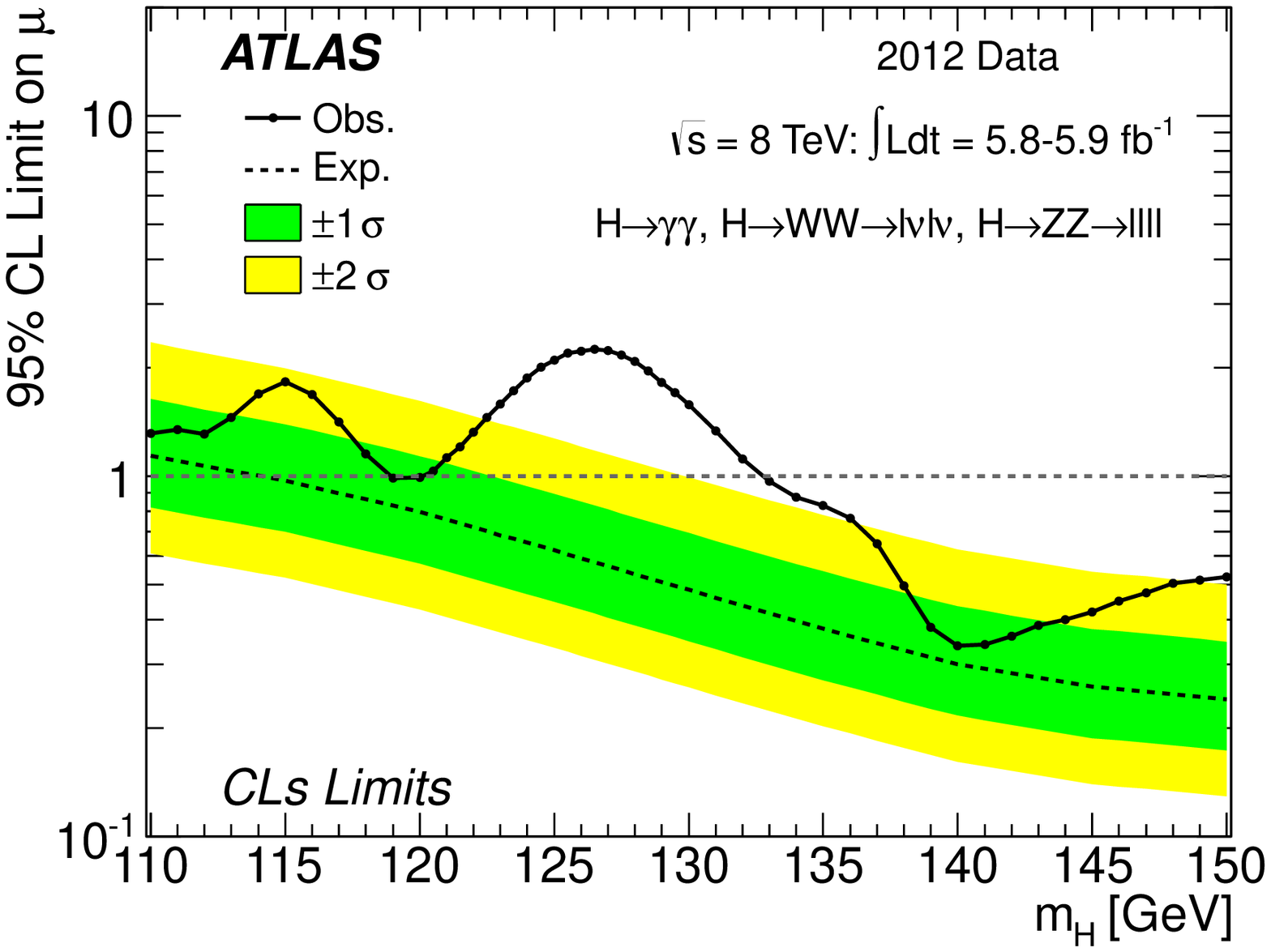}
\includegraphics[width=7.5cm]{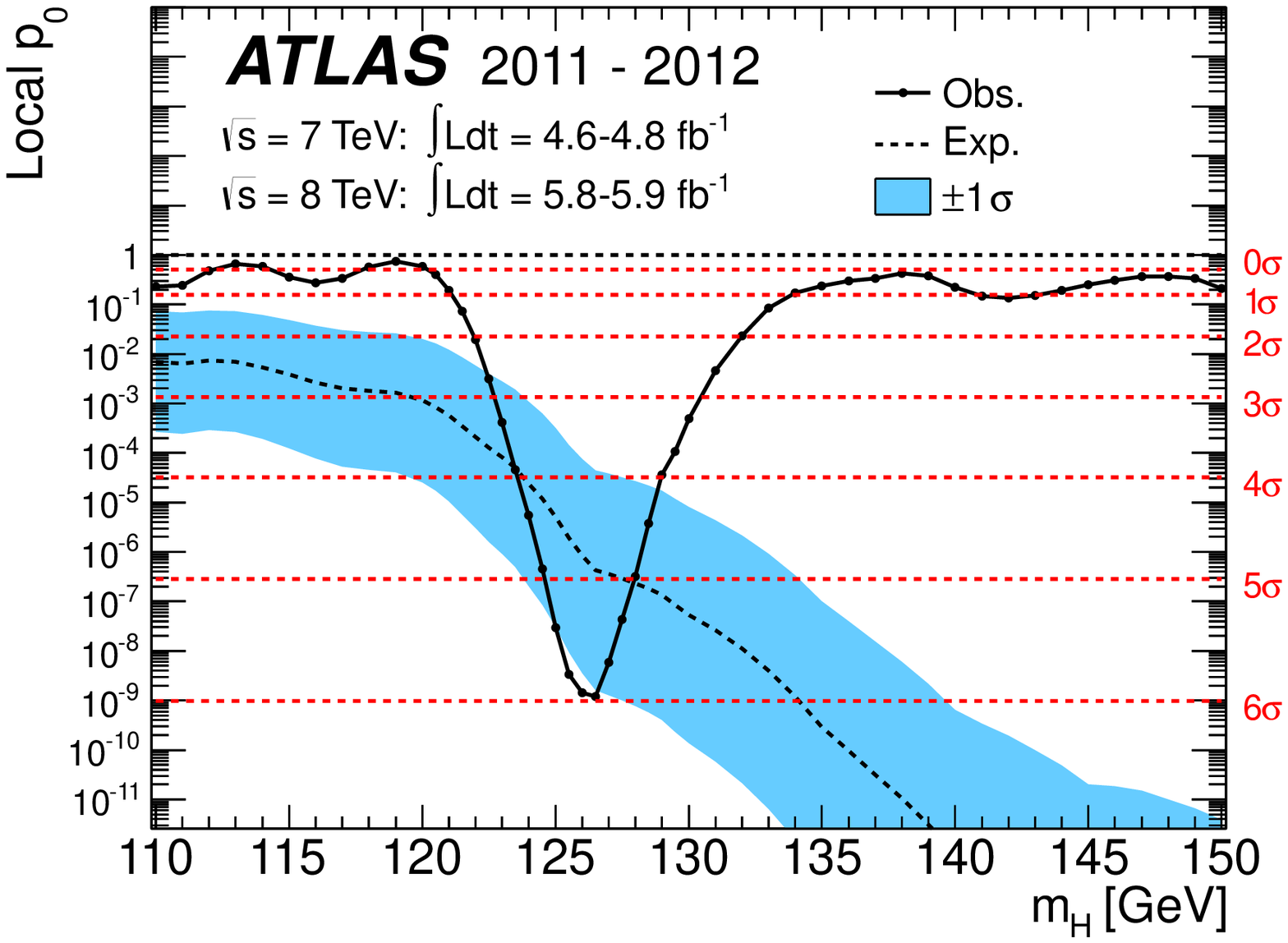}
\includegraphics[width=7.5cm]{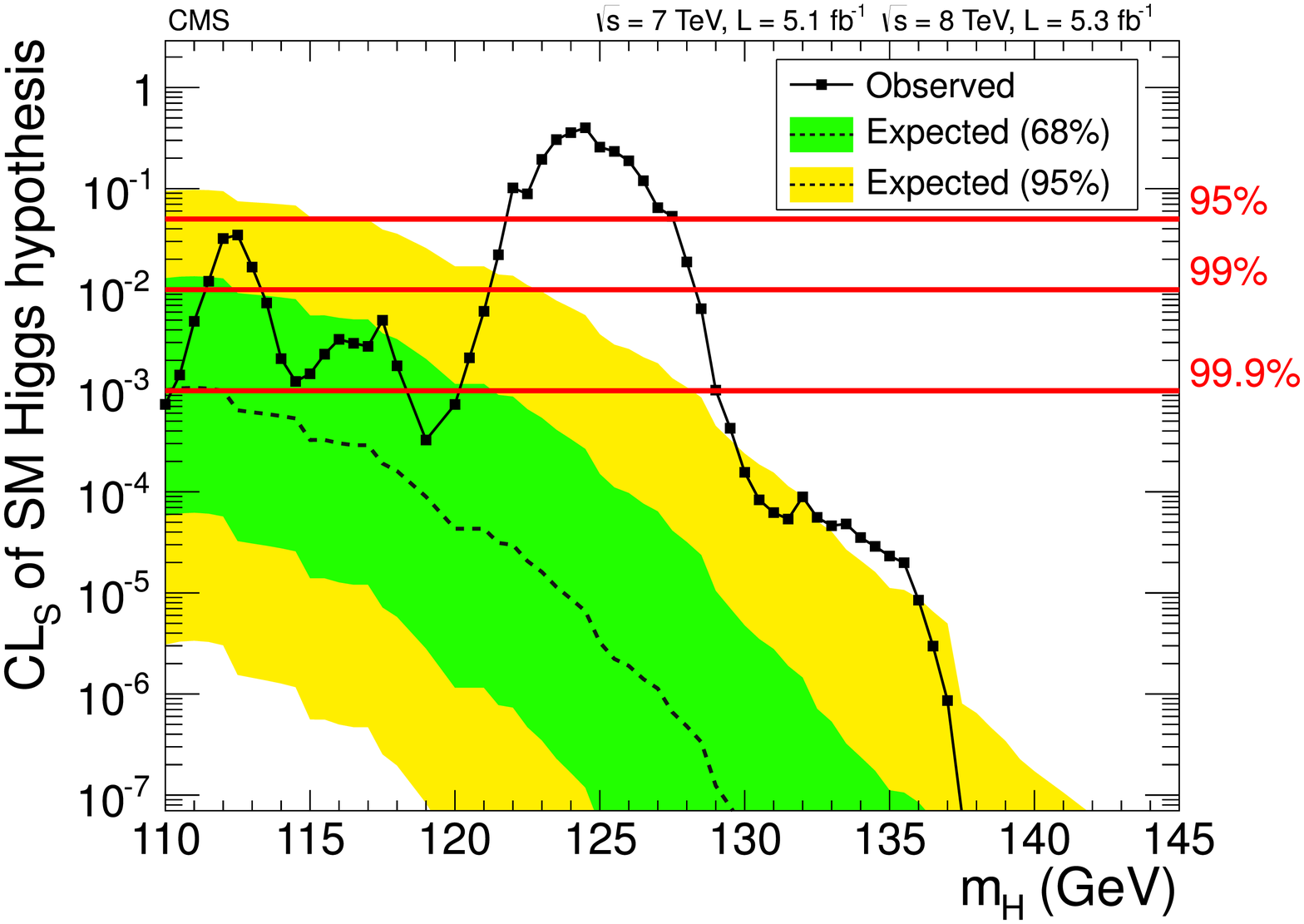}
\includegraphics[width=7.5cm]{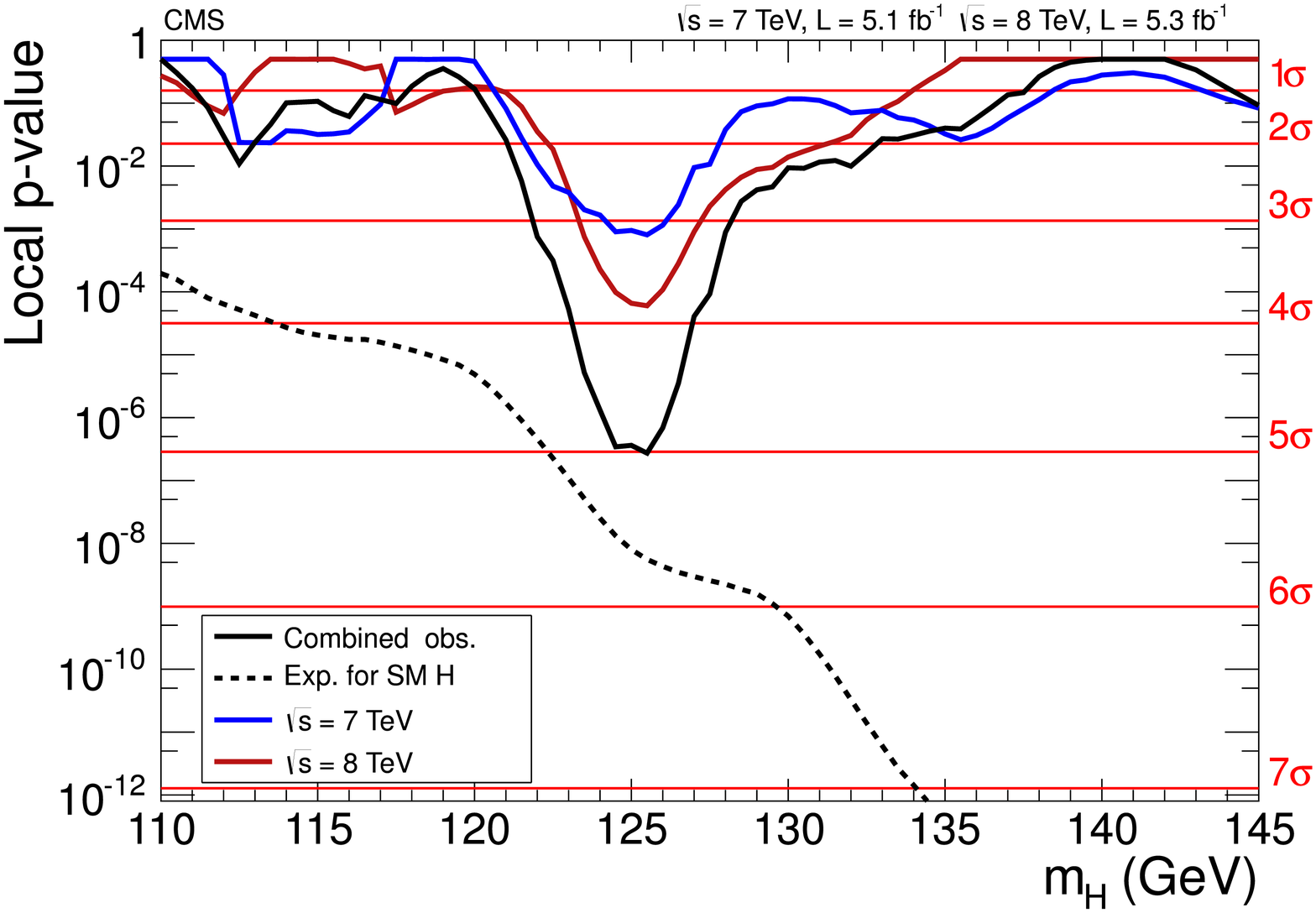}
\caption{ 
The expected distributions and observed values for the exclusion (left) and the observation significances (right)
in searches for the SM Higgs boson
presented by the ATLAS\cite{bib:ATLASfourthJuly} (top) and CMS\cite{bib:CMSfourthJuly} (bottom)
collaborations on July 4th, 2012.  
Both experiments find a significant excess around 125 GeV.
}
\label{fig:fourthJuly}
\end{center}
\end{figure}

To conclude the discussion on hypothesis testing, there is certainly no better illustration 
than Figure~\ref{fig:fourthJuly},
taken from the results announced by the ATLAS and CMS experiments on July 4th, 2012: in the context of the search
for the SM Higgs boson, both collaborations 
established the observation of a new particle with a mass around 125 GeV.
\section{Conclusions}
These lectures aimed at providing a pedagogical overview of  probability and statistics. The choice
of topics was driven by practical considerations, based on the tools and concepts actually encountered
in experimental high-energy physics. A bias in the choices, induced by the author's perspective and personal experience 
is not to be excluded. The discussion was completed with  examples from recent results, mostly (although
not exclusively)
stemming from the $B$-factories and the LHC experiments. 
\section{Acknowledgements}
I would like to express my gratitude to all who helped make the AEPSHEP  2012 school a success:
the organizers, the students, my fellow lecturers and discussion leaders. 
The highly stimulating atmosphere brought fruitful interactions during lectures, discussion
sessions and conviviality activities.


\begin{thebibliography}{99}

\bibitem{bib:Kendall}  A.~Stuart, J.K.~Ord, and S.~Arnold,
{\it Kendall's Advanced Theory of Statistics}, Vol. 2A: {\it Classical
Inference and the Linear Model} 6th Ed., Oxford Univ. Press (1999),
and earlier editions by Kendall and Stuart.

\bibitem{bib:Barlow} R.J.~Barlow, {\it Statistics: A Guide to the Use of
Statistical Methods in the Physical Sciences}, Wiley, 1989.

\bibitem{bib:Cowan} G.D.~Cowan, {\it Statistical Data Analysis},
Oxford University Press, 1998.

\bibitem{bib:James} F.~James, {\it Statistical Methods in Experimental
Physics}, 2nd ed., World Scientific, 2006.

\bibitem{bib:Lyons} L.~Lyons, {\it Statistics for Nuclear and Particle
Physicists}, Cambridge University Press, 1986.

\bibitem{bib:PDG} 
  J.~Beringer {\it et al.}  [Particle Data Group Collaboration],
  Phys.\ Rev.\ D {\bf 86}, 010001 (2012).

\bibitem{bib:ROOT} 
  R.~Brun and F.~Rademakers,
  Nucl.\ Instrum.\ Meth.\ A {\bf 389}, 81 (1997).

\bibitem{bib:RooFit} 
  W.~Verkerke and D.~P.~Kirkby,
  eConf C {\bf 0303241}, MOLT007 (2003).

\bibitem{bib:ATLASmumu} 
  The ATLAS collaboration,
  ATLAS-CONF-2013-088.

\bibitem{bib:gFitter} 
  M.~Baak, M.~Goebel, J.~Haller, A.~Hoecker, D.~Kennedy, R.~Kogler, K.~Moenig and M.~Schott {\it et al.},
  Eur.\ Phys.\ J.\ C {\bf 72}, 2205 (2012).

\bibitem{bib:alphaBaBar} 
  J.~P.~Lees [BaBar Collaboration],
  Phys.\ Rev.\ D {\bf 87}, no. 5, 052009 (2013).

\bibitem{bib:KspipiBaBar}
  B.~Aubert {\it et al.}  [BaBar Collaboration],
  Phys.\ Rev.\ D {\bf 80} (2009) 112001.

\bibitem{bib:KspipiBelle} 
  J.~Dalseno {\it et al.}  [Belle Collaboration],
  Phys.\ Rev.\ D {\bf 79}, 072004 (2009).


\bibitem{bib:TMVA}
        A.~Hoecker, P.~Speckmayer, J.~Stelzer, 
        J.~Therhaag, E.~von Toerne, and H.~Voss,
        ``TMVA: Toolkit for Multivariate Data Analysis,''
        PoS A CAT 040 (2007).

\bibitem{bib:HistFactory} 
  K.~Cranmer {\it et al.}  [ROOT Collaboration],
  ``HistFactory: A tool for creating statistical models for use with RooFit and RooStats,''
  CERN-OPEN-2012-016.

\bibitem{bib:RooStats} 
  L.~Moneta, K.~Belasco, K.~S.~Cranmer, S.~Kreiss, A.~Lazzaro, D.~Piparo, G.~Schott and W.~Verkerke {\it et al.},
  The RooStats Project,''
  PoS ACAT {\bf 2010}, 057 (2010).

\bibitem{bib:SMHiggsLEP} 
  R.~Barate {\it et al.}  [LEP Working Group for Higgs boson searches and ALEPH and DELPHI and L3 and OPAL Collaborations],
  Phys.\ Lett.\ B {\bf 565}, 61 (2003).

\bibitem{bib:Read} 
  A.~L.~Read,
  J.\ Phys.\ G {\bf 28}, 2693 (2002).

\bibitem{bib:ATLASCMS} 
  [ATLAS and CMS Collaborations],
  ``Procedure for the LHC Higgs boson search combination in summer 2011,''
  ATL-PHYS-PUB-2011-011, CMS-NOTE-2011-005.

\bibitem{bib:asimov} 
  G.~Cowan, K.~Cranmer, E.~Gross and O.~Vitells,
  Eur.\ Phys.\ J.\ C {\bf 71}, 1554 (2011)

\bibitem{bib:ATLASfourthJuly} 
  G.~Aad {\it et al.}  [ATLAS Collaboration],
  Phys.\ Lett.\ B {\bf 716}, 1 (2012).

\bibitem{bib:CMSfourthJuly} 
  S.~Chatrchyan {\it et al.}  [CMS Collaboration],
  Phys.\ Lett.\ B {\bf 716}, 30 (2012).

\end{thebibliography}
\end{document}